\documentclass[journal]{IEEEtran}

\usepackage{graphicx}
\usepackage{amsmath, amssymb}
\usepackage{braket}
\usepackage{booktabs}
\usepackage{threeparttable}
\usepackage{placeins}
\usepackage{tabularx}
\usepackage{comment}
\usepackage{hyperref}
\hypersetup{colorlinks,allcolors=black}
\usepackage{orcidlink}
\usepackage{array}

\ifCLASSINFOpdf
\else
\fi

\begin{document}
\title{Quantum Knowledge Distillation: State-of-the-art, Applications and Open Problems}
\author{\mbox{Aldana Al-Sulaiti\textsuperscript{1}~\orcidlink{0009-0006-8767-5028}}, \mbox{Kuan-Cheng Chen\textsuperscript{1}}, \mbox{Muhammad Bilal Akram Dastagir\textsuperscript{1}~\orcidlink{0000-0003-2990-4604}}, \mbox{Ahmed Farouk\textsuperscript{1}~\orcidlink{0000-0001-8702-7342}}, and \mbox{Saif Al-Kuwari\textsuperscript{1}~\orcidlink{0000-0002-4402-7710}}
\\[4pt]
\small\textsuperscript{1}\textit{Qatar Center for Quantum Computing,
College of Science and Engineering,
Hamad Bin Khalifa University, Doha, Qatar}}

\maketitle

\begin{abstract}
Knowledge distillation transfers predictions, representations, or other supervisory information between learning models and is increasingly investigated in quantum machine learning. Its applications extend beyond compression to cross-architecture transfer, circuit approximation, distributed learning, and classical deployment of models trained using quantum-derived supervision. This survey reviews these approaches using two complementary classification axes: teacher–student configuration and the role of quantum resources in the distillation process. Quantum-inspired methods are discussed separately from methods that simulate quantum computations classically or execute them on quantum hardware. We explain how measurement probabilities, expectation values, classical readouts, state overlaps, and circuit-level objectives support different transfer mechanisms. We then critically assess the presented results against baseline design, execution conditions, uncertainty, and resource costs. The evidence shows that distillation can improve selected quantum students, but benefits are task- and architecture-dependent, and several studies find negligible gains or even performance degradation. Evaluations on physical quantum hardware remain limited, with some testing individual quantum components rather than complete learning pipelines. Heterogeneous protocols and incomplete resource reporting prevent general conclusions about practical quantum advantage. We identify priorities for controlled benchmarking, noise-aware learning, reproducible evaluation, and efficient transfer between quantum and classical models.
\end{abstract}

\begin{IEEEkeywords}
Knowledge Distillation, Quantum Machine Learning, Hybrid quantum–classical Learning, Variational Quantum Circuits, Model Compression, Resource Requirements, Benchmarking.
\end{IEEEkeywords}

\IEEEpeerreviewmaketitle

\section{Introduction}
\IEEEPARstart{T}{he} rapid advancement of deep learning has enabled increasingly capable models for computer vision, natural language processing, including large language models (LLMs), and medical diagnostics. 
However, their computational and memory requirements can limit deployment on resource-constrained platforms, particularly edge devices. To address these challenges, various model compression techniques have been proposed, including pruning, quantization, low-rank factorization, and knowledge distillation (KD) \cite{gou2021knowledge}. KD trains a student model using supervisory information supplied by a teacher, commonly transferring the predictive behavior of a large model or ensemble to a smaller model~\cite{hinton2015distilling}. Therefore, unlike other compression techniques that directly reduce the model's complexity, KD transfers the teacher's predictive behavior to the student. Classical KD relies on soft-label supervision, with various strategies proposed to improve training stability, performance, and model compression~\cite{gou2021knowledge, hinton2015distilling}. Distillation can improve student performance or reduce deployment requirements, but its effectiveness depends on the teacher, student architecture, training objective, and evaluation settings \cite{gou2021knowledge}. These capabilities motivate the investigation of KD in Quantum Machine Learning (QML), where differences in model representations and execution requirements introduce additional challenges for knowledge transfer.

In parallel, QML investigates the use of quantum computations within learning models and algorithms. However, noisy intermediate-scale quantum (NISQ) hardware imposes constraints through limited qubit counts, gate errors, and decoherence \cite{bharti2022noisy}. 
Therefore, quantum KD extends classical KD by enabling the transfer of supervisory information between classical, quantum, and hybrid models and, in some settings, training smaller quantum circuits. It can support compressing quantum circuits into more hardware-efficient models and facilitate knowledge transfer across different classical and quantum architectures. A smaller quantum parameterization, however, does not necessarily imply lower end-to-end computational cost: state preparation, repeated measurements, gradient estimation, and classical preprocessing also contribute to execution requirements.

Beyond model compression, quantum KD can address the challenges of trainability by providing additional supervisory signals during training. Certain parameterized quantum models exhibit barren plateaus, in which cost-function gradients concentrate near zero with variance decreasing exponentially with qubit number \cite{mcclean2018barren}. Their occurrence depends on factors including circuit architecture, initialization, cost function, and noise. However, improved convergence in a particular experiment does not establish that distillation prevents barren plateaus; such a claim requires explicit analysis of gradient behavior and scaling \cite{li2024hybrid, nishan2026quantummedkd}.

Research on quantum KD remains fragmented across teacher–student configurations, transfer mechanisms, and application domains. Differences in experimental protocols, baseline selection, and resource reporting make it difficult to compare findings and determine when distillation offers a meaningful benefit. This survey organizes the literature around two complementary axes: the teacher–student configuration and the role of quantum resources in the distillation process. It examines quantum and hybrid methods alongside quantum-assisted distillation, while treating quantum-inspired methods separately from approaches involving simulated or physical quantum computation. Within this framework, methodological choices, experimental evidence, and reported resource requirements are compared to identify supported findings and unresolved questions.

Existing surveys cover various aspects of KD, including types of knowledge, distillation schemes, and algorithms \cite{gou2021knowledge, moslemi2024survey, mansourian2025comprehensive}. Some of these surveys also compare the performance of existing methods \cite{gou2021knowledge, mansourian2025comprehensive}. However, they primarily focus on KD of classical models and do not consider quantum or hybrid models as either teachers or students. In parallel, reviews of quantum transfer learning discuss how a pre-trained classical or quantum model can be reused to train hybrid models \cite{mosley2024quantum, sun2026research}, but they do not address distillation objectives or the teacher-student training framework. Table~\ref{tab:survey_comparison} shows the comparison of existing KD-related surveys with our work. In contrast, this review focuses on recent advancements in quantum KD. It introduces classical KD as a foundation for its quantum extension and covers distillation across different teacher-student configurations, including quantum-enhanced and quantum-inspired distillation, their applications, and benchmarking.

\begin{table*}[t]
	\centering
	\caption{Comparison with existing knowledge distillation surveys.}
	\label{tab:survey_comparison}
    	\begin{tabular}{ccccccc}
	\toprule
	\textbf{Ref.} & \textbf{Tutorial} & \textbf{Classical KD} & \textbf{Quantum KD} & \textbf{Quantum-inspired} & \textbf{Applications} & \textbf{Benchmarking}\\
\midrule
         \cite{gou2021knowledge} & $\times$ & $\checkmark$ & $\times$ & $\times$ & $\checkmark$ & $\checkmark$ \\
        \cite{moslemi2024survey} & $\times$ & $\checkmark$ & $\times$ & $\times$ & $\checkmark$ & $\circ$ \\
         \cite{mansourian2025comprehensive} & $\times$ &  $\checkmark$ & $\times$ & $\times$ & $\checkmark$ & $\checkmark$ \\
        This work & $\checkmark$ & $\circ$ & $\checkmark$ & $\checkmark$ & $\checkmark$ & $\checkmark$ \\
\bottomrule
\multicolumn{7}{c}{$\checkmark$ = covered, $\times$ = not covered, $\circ$ = partially covered. Quantum KD includes C2Q, Q2Q, Q2C, and quantum-enhanced KD.}
\end{tabular}
\end{table*}

This survey covers studies on knowledge distillation involving quantum or hybrid quantum-classical models, or quantum resources incorporated into the distillation process. The review covers classical-to-quantum, quantum-to-quantum, quantum-to-classical, quantum-enhanced distillation, and quantum-inspired methods. The analysis focuses on distillation mechanisms, model architectures, applications, experimental evidence, and resource requirements. Related approaches such as quantum transfer learning and quantum-state distillation are distinct from knowledge distillation and are not typically included (unless they explicitly involve a teacher-student distillation objective).

\subsection{Contributions}
This paper provides a survey and tutorial on Quantum Knowledge Distillation. Specifically:
\begin{enumerate}
\item We present the foundations of classical KD, and explain their extension to quantum and hybrid models, including knowledge transfer, training schemes, loss functions, and the construction of distillation targets from quantum outputs.
\item We provide a tutorial on how quantum model outputs, such as measurement probabilities and expectation values, are converted into soft targets for KD.
\item We propose a multidimensional taxonomy for quantum KD, classifying studies by teacher-student configuration (classical-to-quantum, quantum-to-quantum, and quantum-to-classical), quantum model type, transferred knowledge, and optimization goal, with separate categories for quantum-enhanced and quantum-inspired KD.
\item We discuss the open challenges in quantum KD and identify future research directions beyond the existing literature.
\end{enumerate}

\subsection{Organization}
The paper is organized as follows: Section~\ref{sec:kd} introduces the foundations of knowledge distillation, including its main types, loss formulation, and training schemes. Section~\ref{sec:qkd} presents quantum knowledge distillation, including teacher-student configurations and quantum model outputs. Section~\ref{sec:survey} reviews the existing literature according to the proposed taxonomy, covering classical-to-quantum, quantum-to-quantum, quantum-to-classical, quantum-enhanced, and quantum-inspired distillation. 
Section~\ref{sec:application} discusses the application domains of quantum KD, while Section~\ref{sec:benchmarking} examines benchmarking practices across the surveyed studies. Open challenges and future research directions are discussed in Section~\ref{sec:challenges}, and the paper is concluded in Section~\ref{sec:conclusion}.

\section{Knowledge Distillation}
\label{sec:kd}
\subsection{Overview}
KD is a widely studied model compression technique that enables the transfer of knowledge from a high-capacity, large teacher model to a smaller student model \cite{hinton2015distilling}. The main goal is to train a compact model that approximates the teacher model's predictions while maintaining high performance. This is achieved by leveraging both ground-truth labels and the teacher's soft targets. This additional information, gained from the teacher's soft targets (referred to as ``dark knowledge'' \cite{hinton2015distilling}), is provided by the teacher, allowing it to provide more information about inter-class probabilities beyond what hard labels alone can provide. Subsequent studies have extended KD beyond compression to improve generalization \cite{sarfraz2021knowledge}, support federated learning \cite{wu2022communication}, and enable transfer between heterogeneous architectures \cite{binici2024generalizing}. KD methods can be organized along two complementary dimensions: the knowledge transferred, commonly classified as response-based, feature-based, or relation-based, and the training scheme, commonly described as offline, online, or self-distillation.

\subsection{Types of Knowledge}
The knowledge transferred through KD is commonly grouped into three categories \cite{gou2021knowledge}, as illustrated in Fig.~\ref{fig:kd_types}:
\begin{figure*}[t]
    \centering
    \includegraphics[width=\textwidth]{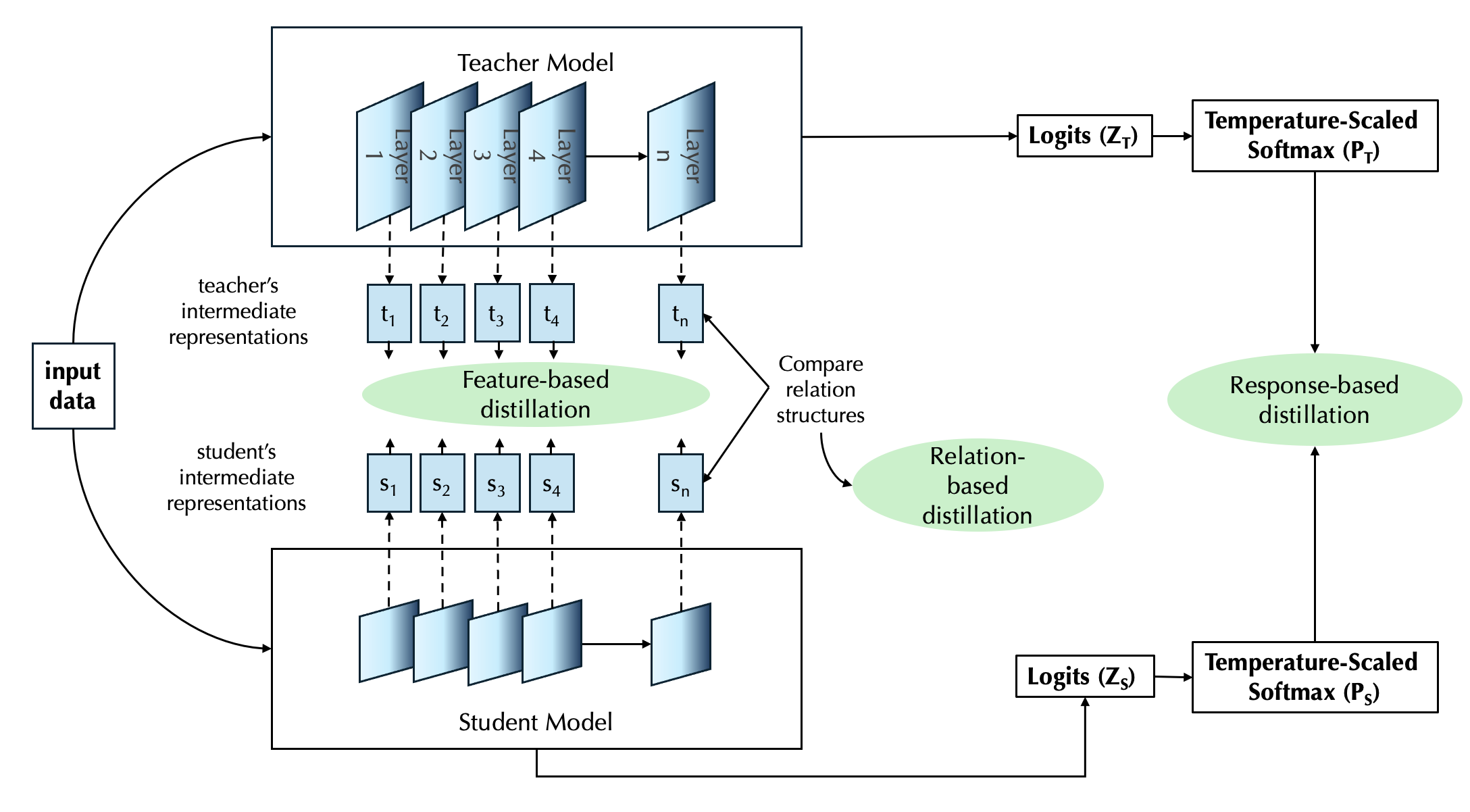}
    \caption{Illustration of the main types of classical KD: response-based, which transfers the teacher's final output distributions; feature-based, which aligns the teacher and student intermediate representations; and relation-based, which aligns the relations computed from teacher and student representations.}
    \label{fig:kd_types}
\end{figure*}
\begin{enumerate}
    \item \textbf{Response-based distillation:} The student is trained to reproduce the teacher’s outputs. In this setting, the student learns from the teacher's softened probability distribution, using a temperature parameter, $\tau$: 
    \begin{equation}
    \label{eq:tempsoftmax}
    p_{M,i}^{(\tau)} = \frac{\exp(z_{M,i}/\tau)} {\sum_{j=1}^{C} \exp(z_{M,j}/\tau)}, \qquad M\in\{T,S\}.
    \end{equation}

    The student captures additional information beyond ground-truth labels using the softened distribution. This additional information is referred to as ``dark knowledge''. The same $\tau$ is applied to both the teacher and student logits before applying softmax, producing softened probability distributions. When $\tau=1$, the softmax reduces to the standard probability distribution used in ordinary classification. In KD settings, $\tau$ is often set greater than $1$ to soften the distribution and reveal relative class similarities. Lower $\tau$ values produce sharp distributions that strongly favor a single class, resembling hard labels, while higher $\tau$ values spread the probability more evenly across classes. However, if $\tau$ is set too high, the distributions may become nearly uniform, diminishing the effectiveness of the distillation signal.
    The student is then trained to minimize a distillation loss based on a divergence measure, specifically the Kullback-Leibler (KL) divergence, between the teacher and student models' softened output distributions. 
    \begin{equation}
    \mathcal{L}_{resp} = \sum_{i=1}^{C}
    p_{T,i}^{(\tau)} \log \left(\frac{p_{T,i}^{(\tau)}}{p_{S,i}^{(\tau)}}\right)
    \end{equation}
    where $p_{T,i}^{(\tau)}$ and $p_{S,i}^{(\tau)}$ are the temperature-scaled teacher and student probabilities for class $i$, respectively.

    Response-based supervision is also used to distill transformer and language models, often in combination with objectives that align intermediate representations  \cite{sanh2019distilbert,jiao2020tinybert}.

    \item \textbf{Feature-based distillation:} It aligns intermediate representations of the teacher and student, such as feature maps or attention maps. These representations provide supervision beyond the final predictions and can guide the student’s internal feature learning. The motivation behind this type of distillation is that the hidden layers of a deep teacher model encode richer semantic information than the final logits alone \cite{romero2014fitnets}. 
    
    FitNets \cite{romero2014fitnets} introduced a prominent approach in which intermediate teacher representations serve as hints for training a student. In this framework, the student is encouraged to match its feature activations with those of the teacher through a feature-matching loss:
    \begin{equation}
    \mathcal{L}_{feat} = \left\lVert F_T(x) - g(F_S(x)) \right\rVert_2^2,
    \end{equation}
    where $F_T(x)$ and $F_S(x)$ denote the teacher and student intermediate feature representations for input $x$ respectively, which may be vectors or multi-dimensional feature maps. The mapping $g(\cdot)$ is used as a projection function to accommodate differences in the feature dimensions between the teacher and student models, including spatial and channel dimensions. This projection is important because it handles feature-dimensionality mismatches between the teacher and student models, especially when their architectures differ in depth or width. For tensor-valued feature maps, the squared Euclidean norm can be evaluated after flattening, while any feature normalization should be specified separately. In quantum or hybrid models, features used for alignment may include classical activations or estimated expectation values of quantum observables; access to the full quantum state is not generally assumed.

    Extensions of feature-based KD transfer attention maps \cite{zagoruyko2016paying}, channel correlations \cite{gou2022multilevel, gou2022channel}, representations from multiple stages \cite{xu2022multiple}, and information derived through cross-attention mechanisms \cite{hsu2024csakd}. These approaches provide several ways to align internal representations when teacher and student architectures differ. Their application to quantum and hybrid students is discussed in Section~\ref{subsec:c2q}.
    
    \item \textbf{Relation/Similarity-based distillation:} Relation-based KD transfers structural relationships among data samples or feature embeddings rather than matching final outputs or intermediate activations \cite{park2019relational}. The main idea is that relational information is encoded in the teacher model's representation space, capturing the relative structure of how samples are organized with respect to one another, even when architectures differ.
    An example is relational knowledge distillation (RKD), introduced by Park et al. \cite{park2019relational}, in which the student is encouraged to match the distance- and angle-wise relations between data samples through a loss function. The distance-wise relation compares pairwise distances between samples in the teacher's representation space to their corresponding distances in the student representation space. A simplified distance-wise loss can be written as:
    \begin{equation}
    \mathcal{L}_{dist} = \sum_{i,j} \left( d_{ij}^{T} - d_{ij}^{S} \right)^2,
    \end{equation}
    where $d_{ij}^{T}$ and $d_{ij}^{S}$ denote the pairwise distance between the samples $i$ and $j$ in the teacher and the student representation space respectively. This distillation type is relevant to quantum-enhanced distillation, where inter-sample relationships are computed in a high-dimensional Hilbert space rather than computing classical pairwise distances as discussed in Section~\ref{subsec:qenhanced}. 
    \end{enumerate}

Overall, although KD was first introduced as a compression-based approach primarily focused on matching the teacher and student's final output distributions, subsequent studies have expanded the field by exploring additional forms of transferable knowledge that extend foundational response-based KD, including feature representations and relational structures between samples. 

\subsection{Distillation Objective}
\label{distloss}
In supervised KD settings, the student is trained with a combined objective that balances a task loss on ground-truth labels and a distillation loss that leverages teacher-guided knowledge \cite{gou2021knowledge, mansourian2025comprehensive}. The supervised task loss encourages the student to predict class labels, while the distillation loss encourages the student to approximate the teacher's knowledge. The distillation loss depends on the type of knowledge being transferred, such as response-based, feature-based, or relation-based distillation. For classification tasks, the cross-entropy task loss is defined as:
\begin{equation}
\mathcal{L}_{task} = - \sum_{i=1}^{C} y_i \log(p_i)
\end{equation}
where $C$ is the number of classes, $y_i$ is the ground truth label, and $p_i$ is the predicted probability for class $i$, computed without temperature scaling. 

For response-based KD, the distillation loss is often defined using the Kullback-Leibler (KL) divergence between the teacher and the student softened output distributions \cite{hinton2015distilling, gou2021knowledge}. The KL divergence between two discrete probability distributions, $P$ and $Q$, is defined as:
\begin{equation}
\mathrm{KL}(P \parallel Q) =
\sum_{i=1}^{C}
P(i)
\log
\left(
\frac{P(i)}{Q(i)}
\right).
\end{equation}
In this distillation setting, $P$ denotes the teacher's softened output distribution, $Q$ the student's output distribution, and $C$ the number of classes.
The KL divergence quantifies the discrepancy between the teacher and student output distributions, effectively measuring how well the student approximates the teacher. Larger differences result in higher penalties, guiding the student model to better approximate the teacher's predictive behavior.

The overall student objective is formed by combining the supervised task loss with the distillation loss, depending on the type of knowledge being transferred:
\begin{equation}
\mathcal{L}_{total} = (1-\alpha)\mathcal{L}_{task} + \alpha\mathcal{L}_{distill}
\end{equation}

where $\mathcal{L}_{distill}$ denotes the distillation loss, computed depending on whether the KD is response-based, feature-based, or relation-based. $\mathcal{L}_{task}$ is computed as the supervised task loss with respect to the ground-truth labels. The parameter $\alpha$ balances the trade-off between distillation loss and supervised task loss. A smaller $\alpha$ emphasizes ground-truth labels and prioritizes the supervised task loss, whereas a larger $\alpha$ increases the distillation loss's influence on the model. Setting $\alpha=0$ reduces the objective to standard supervised training with no distillation signal, whereas setting $\alpha=1$ removes the ground-truth supervision, training the student only on the distillation loss. 

For response-based KD with temperature-scaled softmax, the distillation term is commonly scaled by $\tau^2$:
\begin{equation}
\mathcal{L}_{total} = (1-\alpha)\,\mathcal{L}_{task} + \alpha\,\tau^{2}\,\mathcal{L}_{resp}
\label{eq:total}
\end{equation}
The $\tau^2$ term compensates for the reduction in gradient magnitude introduced by temperature scaling in (Eq.~\ref{eq:tempsoftmax}). When the teacher and student logits are divided by $\tau$ before applying softmax, temperature scaling reduces the gradient contribution of the softened KL loss, especially at high temperatures, where the softened teacher and student distributions approach uniformity. Multiplying the distillation term by $\tau^2$ can preserve the contribution of the distillation loss when it is combined with the task loss, so it remains comparable to the task loss \cite{hinton2015distilling}.

In label-free settings, such as unsupervised or self-supervised learning, the supervised task loss may be omitted from the formula, and the student is trained only on the teacher-guided distillation loss \cite{singh2024simple}.

\subsection{Training Schemes}
KD training schemes are categorized based on whether the teacher is fixed prior to the student training. The most common approach is offline distillation, in which the teacher model is first trained independently and then frozen \cite{hinton2015distilling}. The trained teacher is then used to generate soft targets as softened probability distributions to guide the student as a static supervisory signal. This training scheme is particularly prevalent in quantum KD settings, where a pre-trained classical teacher is most commonly frozen and later used to guide the student model \cite{hasan2023bridging, nishan2026quantummedkd}.

The general process of offline KD can be summarized as follows:
\begin{enumerate}
    \item The teacher model is first trained to convergence, and its output logits or probability distribution are saved.
    \item The output logits or probability distribution of the teacher model are softened by a temperature parameter, producing a more informative class distribution referred to as soft targets.
    \item The student model is trained using both hard targets (ground truth labels) and soft targets to minimize a combined loss.
\end{enumerate}

In contrast, online distillation does not involve a pre-trained teacher model; instead, the teacher and student models are trained simultaneously \cite{zhang2018deep}. Therefore, knowledge is dynamically exchanged between the models during training. This scheme is often used in a federated learning setting where no pre-trained central model is available, allowing the student models to benefit from each other's predictions \cite{wu2022communication}. 
Self-distillation is a technique in which a single model improves its own performance by leveraging its intermediate representations or earlier predictions as a form of guidance \cite{zhang2019your, furlanello2018born, gou2021knowledge}. In this scheme, the distinction between the teacher and the student is implicit, as knowledge is transferred within the same model.

\section{Quantum Knowledge Distillation}
\label{sec:qkd}

\subsection{Overview and Teacher-Student Configurations}
\label{sec:overview}
\begin{figure}[t]
    \centering
    \includegraphics[width=\columnwidth]{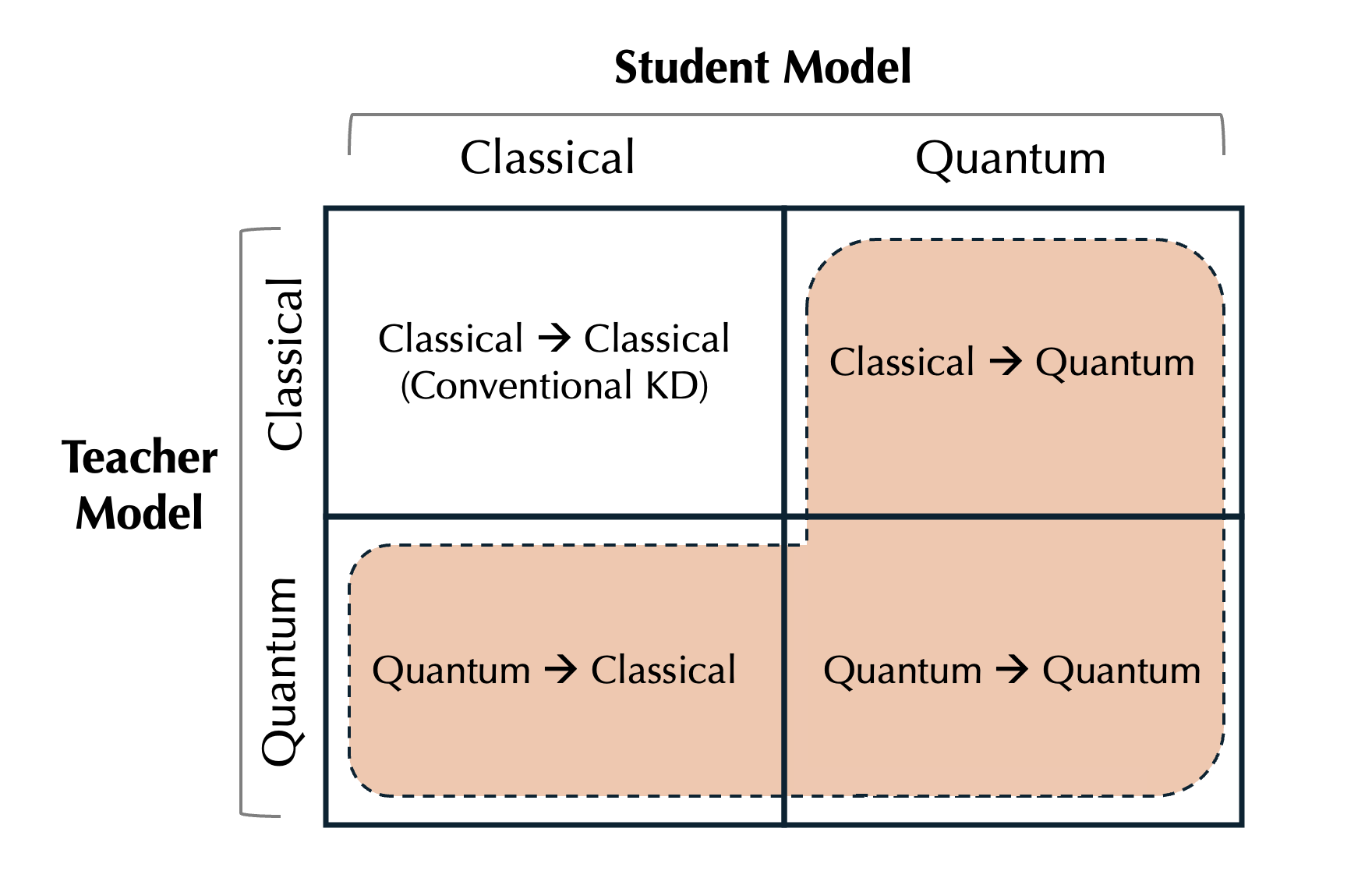}
    \caption{Combinations of teacher-student configurations in quantum knowledge distillation.}
    \label{fig:qkd_taxonomy}
\end{figure}

While knowledge distillation is well established in machine learning, its extension to quantum and hybrid quantum-classical models is relatively new and introduces new challenges and motivations. Quantum KD is generally defined as a knowledge distillation framework in which the teacher, the student, or both involve quantum or hybrid quantum-classical models. This survey focuses on the teacher-student configurations, with quantum-enhanced and quantum-inspired approaches serving as complementary approaches. Depending on the framework, the transferred knowledge can be represented through measurement distributions, expectation values, intermediate features, relational structures, or unitary transformations. Therefore, it is essential to categorize it based on the direction of knowledge transfer. In general, the teacher-student interaction results in four distinct combinations, as shown in Fig.~\ref{fig:qkd_taxonomy}: classical-to-classical (C2C), classical-to-quantum (C2Q), quantum-to-quantum (Q2Q), and quantum-to-classical (Q2C) KD. Classical-to-classical distillation corresponds to conventional KD. Quantum-enhanced and quantum-inspired approaches retain this configuration and are discussed separately as a complementary category. The three combinations involve at least one quantum or hybrid quantum-classical model and represent the teacher-student configuration directions considered in this survey. Table~\ref{tab:qkd_directions} summarizes the directions that follow from these configurations, structuring both the framework presented and the literature review presented in Section~\ref{sec:survey}. Separately, quantum-enhanced distillation is a complementary category in which quantum resources are used within the distillation mechanism, while the teacher and the student are classical networks. This includes utilizing quantum kernels, circuits, or hardware to enhance representation learning or similarity estimation during training. In addition, quantum-inspired KD, reviewed in Section~\ref{subsec:qinspired}, is treated as a separate category, in which classical models use quantum-derived constructions without evaluating any quantum state or circuit.

\begin{table*}[t]
\centering
\caption{Taxonomy of quantum knowledge distillation: teacher-student configuration directions and the complementary quantum-enhanced category}
\label{tab:qkd_directions}
\small
\begin{tabular}{p{0.17\textwidth} 
p{0.26\textwidth} 
>{\raggedright\arraybackslash}p{0.22\textwidth} 
>{\raggedright\arraybackslash}p{0.06\textwidth} 
>{\raggedright\arraybackslash}p{0.06\textwidth} 
>{\raggedright\arraybackslash}p{0.09\textwidth}}
\toprule
Category & Primary Goal & Representative Work & Studies & Hardware & Classical Comparison \\
\midrule

Classical-to-Quantum (C2Q) 
& Train or improve quantum student models using classical
teacher knowledge 
& \cite{hasan2023bridging, piperno2025classical, piperno2024quantum,
li2024hybrid, ovi2025bridging, nishan2026quantummedkd,
li2025quantum, liu2025federated, barbato2026quantum, lazaro2026staged, innan2026towards, amin2025quantum, franco2026foundation, aizpurua2026quantum, roh2024fast, kim2025quantum, roh2025hybrid, boucher2025n}
& 18 & 3 & 10 \\[6pt]

Quantum-to-Quantum (Q2Q) 
& Transfer knowledge between quantum or hybrid quantum-classical models for compression, performance improvement, or efficient deployment. 
& \cite{yan2026distilling, alam2023knowledge, gurung2025llm, krishnamurthy2026early, yu2025knowledge, han2026layerwise, piperno2026knowledge, zhu2025qctkd, oh2025quantum}
& 9 & 2 & 3 \\[6pt]

Quantum-to-Classical (Q2C)
& Transfer quantum-derived supervisory information to a deployable classical student
& \cite{naqvi2026structured, majeed2026compute}
& 2 & 0 & 1 \\[6pt]

Quantum-Enhanced 
& Use quantum resources (kernels, circuits, or hardware) to enhance the distillation process
itself
& \cite{liu2025quantum, chen2026photonic, li2026quantumgated, ni2026quloc} 
& 4 & 1 & 4 \\[6pt]

\hline
\end{tabular}
\begin{tablenotes}
\footnotesize
\item \textit{Studies}: number of surveyed studies in each category. \textit{Hardware}: studies evaluated on a physical quantum processor; emulators are not counted. 
\item \textit{Classical Comparison}: studies that compare against a classical baseline other than the teacher model.
\item \textit{Scope}: the table covers frameworks that use quantum models and quantum computations, whether simulated or executed on real hardware within the teacher, student, or distillation process. Quantum-inspired methods are discussed separately in Section~\ref{subsec:qinspired} and are not included in the table counts.
\end{tablenotes}
\end{table*}

In a classical-to-quantum setting, a classical teacher, usually a strong model, guides a compact quantum student model. In a quantum-to-classical setting, knowledge is transferred from a quantum teacher model to a classical student, enabling the quantum model's learned information to be represented in the classical student. In a quantum-to-quantum setting, both the teacher and the student are either quantum models or hybrid quantum-classical models. Here, knowledge is transferred from a complex, more expressive quantum model to a smaller, more hardware-efficient quantum model. In addition to the above categories, the expressive nature of quantum models has recently led to the emergence of quantum-enhanced KD. In this category, the teacher and the student remain mainly classical, but the distillation process is enhanced through a quantum component \cite{liu2025quantum, chen2026photonic, ni2026quloc, li2026quantumgated}. 

While the primary focus is on the direction of knowledge transfer, quantum KD can also be categorized along several axes. These categories include the type of quantum model utilized (e.g., fully quantum, hybrid quantum-classical, or quantum kernel-based), the form being transferred (e.g., output logits, measurement probabilities, or expectation values), or the primary optimization goal (e.g., performance improvement, circuit compression, or efficient deployment). Table~\ref{tab:crossref} provides a cross-reference of the quantum KD and the quantum-enhanced frameworks based on these dimensions.

\subsection{Constructing Soft Targets From Quantum Outputs}
\label{q-softtargets}
As discussed in Section~\ref{distloss}, classical response-based KD commonly applies the temperature-scaled softmax in Eq.~\ref{eq:tempsoftmax} to logits to generate softened class probabilities \cite{xuan2025exploring}. These softened probabilities are then compared using a divergence measure such as KL divergence, assuming the model produces logits that can be transformed into class probabilities.

Purely quantum models do not inherently produce logits in the same way classical neural networks do. Instead, outputs are obtained via quantum measurement, which can yield either expectation values or probability distributions \cite{bharti2022noisy}. However, hybrid quantum-classical models can produce logits via a classical readout layer after the quantum circuit.

\begin{figure*}[t]
\centering
\includegraphics[width=\textwidth]{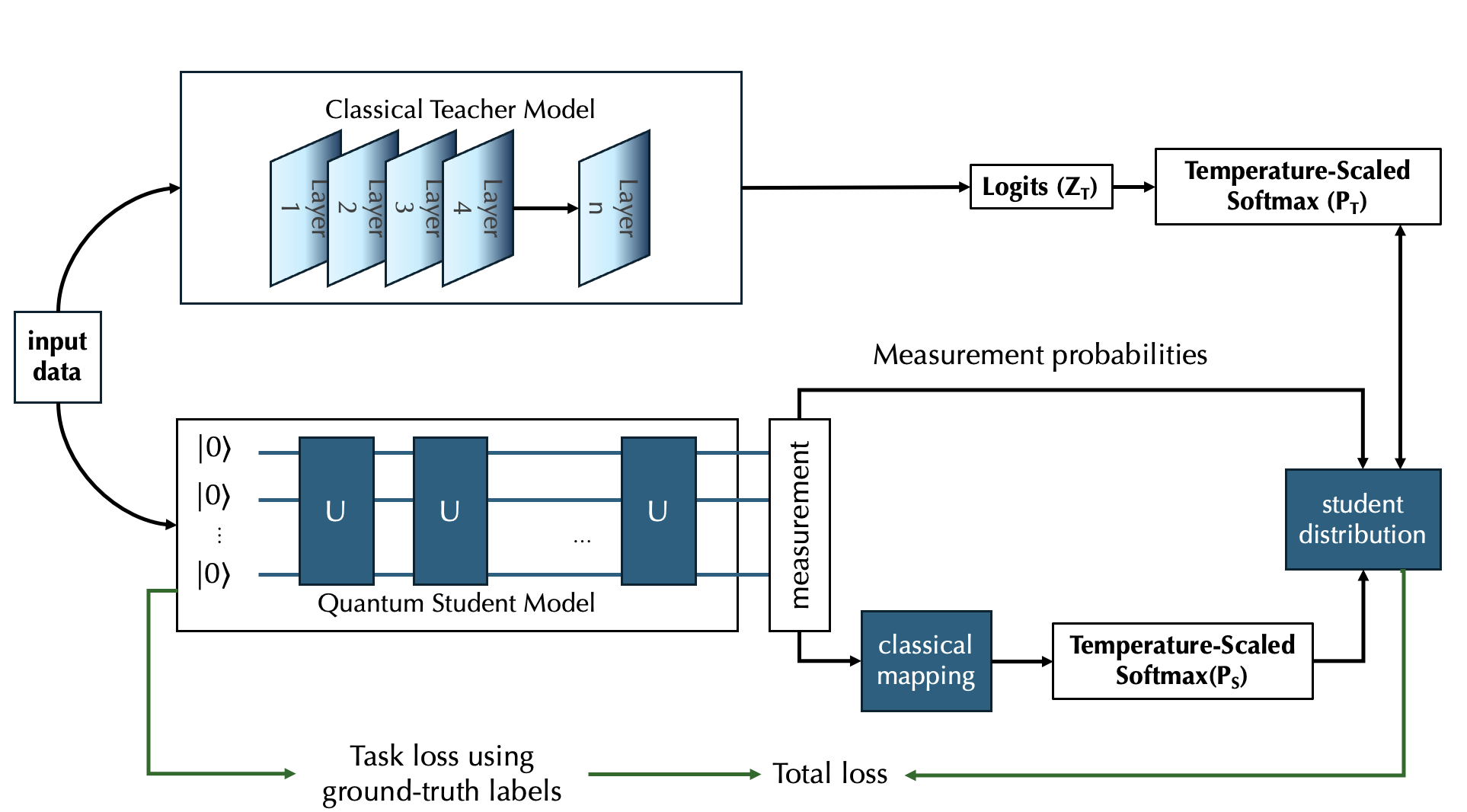}
\caption{The pipeline for classical-to-quantum response-based knowledge distillation involves converting the teacher's logits and the quantum student's outputs into probability distributions for KD, either via measurement probabilities or via a classical readout followed by a temperature-scaled softmax.}
\label{fig:c2q_kd_framework}
\end{figure*}

Each output type requires different treatment before the distillation loss is computed. Given an input $x$, a parameterized quantum circuit prepares a quantum state:
\begin{equation}
|\psi(x;\theta)\rangle = U(x;\theta)|0\rangle^{\otimes n},
\label{eq:state}
\end{equation}
where $U(x;\theta)$ denotes the data-encoding and parameterized quantum circuit applied to an $n$-qubit initial state, and $U(x;\theta)|0\rangle^{\otimes n}$ is the resulting quantum state before measurement.
The appropriate distillation formulation depends on the form of the quantum model output. Three cases arise: measurement probabilities, expectation values, and quantum measurement followed by classical post-processing. Fig.~\ref{fig:c2q_kd_framework} illustrates the choices for handling the output of the classical-to-quantum response-based KD pipeline for distillation.

\textbf{Measurement Probabilities.} If the model output is represented as measurement probabilities, the output is considered a valid probability distribution over measurement outcomes. The probability of observing measurement outcome $k$ is given by Born’s rule:
\begin{equation}
p_{\theta}(k \mid x) = \left| \braket{k|\psi(x;\theta)} \right|^2.
\end{equation}
where $\ket{k}$ denotes a computational basis state and $|\psi(x;\theta)\rangle$ is the quantum state prepared by the quantum circuit.
In a classification setting, measurement outcomes are mapped to class probabilities by assigning one or more measurement outcomes to each class. Although these probabilities sum up to $1$, they correspond to measurement outcomes rather than necessarily to task classes. For classification tasks, one or more outcomes are assigned to each class, and the class probability is obtained by summing the probabilities of the outcomes assigned to it. The resulting class distribution can be compared with the teacher distribution using $\mathrm{KL}(p_T \parallel p_S)$, without temperature scaling, provided both distributions refer to the same classes and the student assigns positive probability wherever the teacher does. For finite-shot estimates, the probability estimator and any smoothing or clipping followed by renormalization must be specified.

\textbf{Expectation Values.} Alternatively, many quantum and hybrid quantum-classical models use expectation values of observables as outputs. Given an observable $O$, the expectation value is defined as:
\begin{equation}
\langle O \rangle_{\theta}
=
\langle \psi(x;\theta) | O | \psi(x;\theta) \rangle.
\end{equation}
Expectation values are measurement-derived scores rather than class probabilities. For example, the expectation value of the Pauli-Z observable is bounded in $[-1, 1]$ and does not generally form a normalized distribution over classes. For a $C$-class task, one approach is to measure $C$ observables and use their expectation values as class scores:
\begin{equation}
o_j = \langle O_j \rangle_{\theta},
\qquad
j \in \{1,\ldots,C\}.
\end{equation}
These scores can be converted into normalized class probabilities using temperature-scaled softmax:
\begin{equation}
p_j^{(\tau)} = \frac{\exp(o_j/\tau)}{\sum_{k=1}^{C}\exp(o_k/\tau)}.
\end{equation}
However, for Pauli observables with \(o_j\in[-1,1]\), the range of probability distributions that can be represented is limited. In particular, the probability assigned to any class is bounded by 
\begin{equation}
    p_j^{(\tau)}\leq\frac{e^{2/\tau}}{e^{2/\tau}+C-1}.
    \label{eq:softmaxbound}
\end{equation}

A suitable classical scale or readout, such as Eq.~\ref{eq:readout}, may therefore be needed to match the teacher’s probability distribution.

In binary classification with a single Pauli-$Z$ expectation value, $\langle Z \rangle_{\theta}\in[-1,1]$, a direct mapping produces two class probabilities without a learned classical readout:
\begin{equation}
p_{\theta}(y=0 \mid x) =
\frac{1+\langle Z \rangle_{\theta}}{2},
\qquad
p_{\theta}(y=1 \mid x) =
\frac{1-\langle Z \rangle_{\theta}}{2}.
\end{equation}

\textbf{Hybrid Quantum-Classical Post-Processing.}
In hybrid quantum-classical settings, measurement probabilities or expectation values can undergo classical post-processing before the distillation or task loss is computed. This post-processing may use a fixed mapping, normalization, or a readout layer to map quantum outputs to class scores. This is particularly relevant when qubits do not yield quantum outputs per class; a classical mapping should be applied to match the correct output dimension before softmax temperature scaling.

Let $\mathbf{o}$ denote a vector of quantum measurement outputs; a classical readout maps these outputs as:
\begin{equation}
\mathbf{z} = W\mathbf{o}+\mathbf{b},
\label{eq:readout}
\end{equation}
where \(\mathbf{z}\in\mathbb{R}^{C}\) contains one class score for each of the $C$ target classes. The resulting class scores can then be converted to a temperature-scaled probability distribution by applying softmax to \(\mathbf{z}\) to produce the student probability distribution, as in conventional KD.

The conventional $\tau^2$ factor is used when both the teacher and student probabilities are produced by the temperature-scaled softmax of the logits. This is because temperature scales the corresponding gradient. This argument does not automatically transfer to a quantum student whose measurement probabilities are given by Born's rule, since no softmax temperature scaling is used for them. In this case, $\tau^2$ may still be used, but should be interpreted as a selected loss weight rather than a gradient correction.
Additionally, a distributional sharpness mismatch may arise when comparing the temperature-scaled teacher distribution with the quantum non-scaled student distribution. Therefore, it is essential to report whether the temperature is applied only to the teacher, to a classical readout of the student, or to both outputs.

\subsection{Distillation in Quantum Settings}
Distillation in quantum settings can operate at different levels of representation. Depending on the framework, the student may be trained to reproduce the teacher's output behavior, intermediate representations, relational structure, or circuit-level operation. The appropriate distillation objective, therefore, depends on the form of knowledge being transferred.

In response-based KD settings, teacher and student predictions must be expressed in compatible forms before computing the loss. For classification tasks, this typically means mapping the predictions to probability distributions over the same task classes, after which an objective such as KL divergence can be applied. In particular, the KL divergence remains applicable once the teacher and student models produce outputs that can be expressed as probability distributions over the same task classes, provided the student assigns positive probability whenever the teacher does. This allows the student model to learn from both ground-truth labels and the teacher model's output behavior. Some work additionally incorporates Jensen-Shannon divergence (JSD) alongside the task loss and KL divergence as a complementary divergence measure \cite{li2025quantum}.

Distillation does not operate only at the output level. For example, Alam et al.~\cite{alam2023knowledge} propose gate-level knowledge distillation, using the Hilbert--Schmidt distance to quantify the discrepancy between the teacher and student unitaries. This extends the notion of what constitutes ``knowledge'' in the quantum KD framework beyond soft targets.

In response-based KD, both the teacher and student models must be represented as probability distributions over the same task classes. Measurement probabilities require an outcome-to-class mapping; expectation values require a conversion or class-score mapping; and hybrid quantum-classical models can produce class logits via a classical post-processing layer after the quantum circuit. In this case, the temperature-scaled softmax is applicable, and the standard KD loss can be applied as in classical neural networks. 

For gradient-based training, the loss gradient is computed by first differentiating the quantum circuit output with respect to its trainable parameters. Consider a circuit with parameters $\theta=(\theta_1,\ldots,\theta_m)$ and an output of this circuit $o(\theta)=\langle O\rangle_{\theta}$, where $O$ is a Hermitian observable. This form covers both output types discussed in Section~\ref{q-softtargets}, since a measurement probability can be written as the expectation value of $\ket{k}\bra{k}$. For a parameter $\theta_l$ appearing in a single gate \(e^{-i\theta_lP/2}\), where the generator \(P\) is Hermitian and \(P^2=I\), the derivative can be obtained using the parameter-shift rule~\cite{schuld2019evaluating} as follows:
\begin{equation}
\frac{\partial o}{\partial \theta_l}
=
\frac{1}{2}
\left[
o\left(\theta+\frac{\pi}{2}\mathbf{e}_l\right)
-
o\left(\theta-\frac{\pi}{2}\mathbf{e}_l\right)
\right],
\label{eq:paramshift}
\end{equation}
where $\mathbf{e}_l$ is the $l$-th unit vector such that only $\theta_l$ is shifted while the other parameters remain fixed. Applying this to each trainable parameter within the circuit yields the gradient of the circuit output.

The distillation loss is computed classically from the circuit outputs $o_j$, either directly when measurement probabilities already define the class distribution, or after a classical readout and softmax when the quantum circuit outputs quantities such as expectation values. Its derivative with respect to a circuit parameter is therefore obtained via the chain rule as:
\begin{equation} \frac{\partial \mathcal{L}}{\partial \theta_l} = \sum_j \frac{\partial \mathcal{L}}{\partial o_j} \frac{\partial o_j}{\partial \theta_l}, \label{eq:chainrule} \end{equation}
where $o_j$ denotes the $j$-th circuit output.  The parameter-shift rule thus evaluates the circuit outputs' changes with respect to a circuit parameter, while the standard backpropagation evaluates how the loss changes with respect to each circuit output. For other generators or parameters shared across gates, gradient-based optimization requires an appropriate differentiation rule. Derivative-free methods can also be used to optimize the circuit.

\textbf{A representative distillation framework.} Figure~\ref{fig:c2q_kd_framework} summarizes a representative classical-to-quantum response-based distillation pipeline followed by many distillation frameworks. The components are combined into a single training loop: a classical teacher is first trained to convergence on the target task and then frozen, so its parameters remain fixed throughout student training. 

For each input \(x\), the teacher produces a logit vector \(\mathbf{z}_T(x)\in\mathbb{R}^{C}\), which is converted into a softened class distribution \(p_T^{(\tau)}(y\mid x)\) by applying temperature-scaled softmax in Eq.~\ref{eq:tempsoftmax}.

In this setting, the same input used for the teacher is encoded into a quantum state, and every VQC block applies parameterized rotation and entangling gates as described in Eq.~\ref{eq:state}. Measurement is then performed, producing either a measurement distribution or a set of expectation values. If the number of measured quantities does not match the number of task classes, a classical readout maps the measurement output to class logits by Eq.~\ref{eq:readout}. If the output is not already a class distribution, the same temperature-scaled softmax used for the teacher is applied to produce the student probability distribution.

The student is then trained using a combined objective in Eq.~\ref{eq:total}. This objective combines task loss, derived from ground-truth labels, and distillation loss, derived from the teacher's softened output distribution. The terms are balanced by \(\alpha\), and the distillation term is commonly scaled by \(\tau^2\) in the case that both teacher and student probabilities are obtained through temperature-scaled softmax; otherwise, the use of \(\tau^2\) should be treated as a selected loss weight as discussed in Section~\ref{q-softtargets}.

In a hybrid student, classical components may occur before the circuit, after the circuit, or both. Their parameters are updated using standard backpropagation. In both fully quantum and hybrid students, the variational circuit parameters can be updated using the parameter-shift rule in Eq.~\ref{eq:paramshift}, which requires additional circuit executions at shifted parameter values. This is a characteristic of hybrid quantum-classical training and contributes substantially to the training cost discussed in Section~\ref{sec:benchmarking}.

\section{Approaches}
\label{sec:survey}
Following the quantum teacher-student configurations introduced in Section~\ref{sec:overview}, this section reviews the existing literature on quantum knowledge distillation according to the teacher-student configuration directions first (C2Q, Q2Q, and Q2C), followed by quantum-enhanced and quantum-inspired KD.

\subsection{Classical-to-Quantum KD (C2Q)}
\label{subsec:c2q}
Classical-to-quantum KD refers to the transfer of knowledge from a classical teacher to a quantum student model. This paradigm is particularly appealing as large classical models, such as deep neural networks and large language models (LLMs), achieve strong predictive performance but require significant computational resources. By distilling the knowledge of large classical models into quantum models, researchers aim to develop more compact learning systems while maintaining high performance.
Research in this area has focused on three main questions: whether knowledge can be effectively transferred from classical to quantum models, how the design of the quantum student influences distillation performance, and whether such approaches can generalize across different application domains.
Some frameworks use quantum circuits as part of a classical pipeline. These frameworks sit at the boundary between C2Q and quantum-enhanced distillation. In these cases, we classify each work by the teacher-student configuration described in its methodology.

A central question in C2Q KD is whether knowledge can be effectively transferred across the classical-quantum boundary. Hasan et al. \cite{hasan2023bridging} proposed a framework in which several classical teacher models, such as LeNet, AlexNet, ResNet, DenseNet, and EfficientNet, are used to distill their knowledge into a compact QNN model. The work reports improved classification accuracy for distilled four- and eight-qubit students in simulation. These results do not establish reduced hardware error-mitigation requirements or lower end-to-end training cost. Experiments across several datasets used in this work demonstrated consistent improvements in accuracy compared to non-distilled quantum models. Through several ablation tests, the authors found that using fully connected layers along with amplitude encoding yielded the best results among the different encoding strategies and dimensionality reduction techniques, both with and without distillation. 

A second study addresses the same question at a smaller scale, where the capacity gap between the teacher and the student is relatively narrower. Piperno et al. \cite{piperno2024quantum} examined four hybrid quantum student architectures under C2Q distillation: a hardware-efficient ansatz, a universal circuit, a universal circuit followed by Hadamard (H) gates, and a universal circuit with mid-circuit measurement, each paired with a classical MLP post-processing layer mapping qubit measurements to class probabilities. The teacher is a classical MLP model trained on a nonlinearly separable XOR dataset. Using a combined $\mathcal{L} = \mathcal{L}_{CE} + \mathcal{L}_{KL}$ objective, this study showed that most quantum students improved with distillation, although one student variant scored lower with distillation than without, and the smallest student uses 74 parameters, compared against a classical baseline student with 195 parameters. However, hybrid architectures exhibit high variance in $F_1$ scores, suggesting sensitivity to parameter initialization. Both \cite{hasan2023bridging} and \cite{piperno2024quantum} establish that the transfer of knowledge from a classical teacher to a quantum student is feasible; the distilled quantum students mostly outperform their non-distilled counterparts. However, the properties that make the transferred signals useful to the student are not isolated, which provides an incentive for further research.

In contrast to the results above, Innan et al. \cite{innan2026towards} provide a controlled evaluation of five quantum transfer learning methods under a shared protocol. One of these methods is ED-QTL, which re-implements the C2Q KD approach of Hasan et al. \cite{hasan2023bridging}. The average logits of three classical teacher models (ResNet18, ResNet34, and DenseNet121) are used to supervise a quantum student using a combination of cross-entropy loss and KL divergence on features extracted from a frozen ResNet18 backbone. On Fashion-MNIST, ED-QTL reaches 85.45\% with four qubits, and 85.88\% with six; the latter falls below the range of the other six-qubit methods (87.15\%--88.28\%). On the Hymenoptera (Ants vs. Bees) dataset, ED-QTL performs worst among the evaluated methods at both qubit counts, reaching 71.24\% and 74.51\% compared with the range of 86.71--94.99\% for the other quantum methods. The authors note that ED-QTL does not translate guidance from the teacher into stronger predictive performance, although it requires substantially less training time.

To investigate how the student's architecture shapes the benefits of distillation, Piperno et al. \cite{piperno2025classical} isolate the impact of hybridization and explore whether the benefits of C2Q KD differ between fully quantum and hybrid architectures. Hybrid models achieve higher absolute $F_1$ scores because the classical MLP component contributes to the final classification task, while fully quantum models consistently outperform their non-distilled baselines. For fully quantum models, predictions are obtained by grouping quantum states into sets and computing cumulative probabilities. In contrast, hybrid models employ a variational quantum circuit followed by a multi-layer perceptron (MLP) for the final classification task.

Li et al. \cite{li2024hybrid} extend C2Q KD to a continuous-variable quantum model. This model is distinct from qubit-based architectures, as data is encoded in the continuous amplitudes of optical modes rather than in discrete two-state qubits. A CNN teacher transfers knowledge to a quantum model through a KL-divergence-based distillation objective. Testing on the MNIST and Fashion-MNIST datasets shows that C2Q KD can also be applied beyond qubit-based architectures. However, increasing the number of modes does not always yield further gains in accuracy, indicating that adding more quantum resources does not necessarily improve performance.

Another design dimension is the entangling structure within the quantum circuit. To bridge the performance gap between QNNs and classical models, Ovi et al. \cite{ovi2025bridging} proposed a C2Q KD framework that accounts for quantum-environment constraints, such as qubit instability and resource limitations. The authors introduced a feature-interaction mechanism based on multiple CNOT entanglements. This architecture enables inter-qubit interactions, allowing it to better capture the complex relationships among features within the quantum circuit. Experimental results on the MNIST dataset demonstrated consistent improvements in accuracy and convergence compared to baseline quantum models.

Beyond the student's architecture, another design choice is which representation the student matches. Barbato et al. \cite{barbato2026quantum} diverge from the prevalent response-oriented C2Q framework by aligning intermediate feature representations instead of final output distributions. The study employs a C2Q framework in which a classical CNN teacher guides a hybrid student with a 5-qubit circuit for binary tasks, and a 4-qubit circuit for multi-class tasks. The training objective combines cross-entropy loss and MSE feature-matching loss rather than using an output-level divergence term. The hybrid student is compared with a parameter-matched classical student on six binary tasks and four multi-class configurations from MNIST and Fashion-MNIST, outperforming the classical student in five of the binary tasks and two of the multi-class configurations. The authors attribute this advantage to the limited expressive capacity of compact classical classifiers at a small parameter budget. Both students are distilled from the same teacher, so the comparison assesses the relative performance of the hybrid and classical student architectures rather than the effect of the distillation itself. On the MNIST 0 versus 9 task, 99.04\% accuracy is reported in simulation and 80.50\% on the IQM Garnet processor, although the hardware results come from a 5-qubit, two-layer circuit retrained in noiseless simulation.

Boucher et al. \cite{boucher2025n} distill a pretrained TinyViT into a quantum vision transformer, aligning intermediate representations rather than the final outputs. The student replaces the linear query, key, and value projections of self-attention with an $n$-qubit PQC, reducing the attention mechanism's parameter scaling from $O(n^2)$ to $O(n)$. The transfer is feature-based, with dimensional mismatch resolved on the teacher's side: TinyViT's classification head is split into two linear layers, with an intermediate width matching the qubit count. MSE aligns the student's measurement-derived representation with this target, and the teacher's final linear layer is then copied into the student.

A further question is whether C2Q KD can extend beyond the conventional image-classification benchmarks. Nishan et al. \cite{nishan2026quantummedkd} evaluated their framework on a real-world clinical dataset for pediatric pneumonia detection, achieving a parameter compression ratio exceeding $10^5$. Three classical teachers (ResNet-50, EfficientNet-B0, and Xception) provide supervisory signals to a PQC student. These studies indicate that the distillation signal alone does not determine the quality of the outcome. The encoding strategy, circuit topology, and entangling structures also affect the magnitude of gains.  

Amin et al. \cite{amin2025quantum} also apply C2Q KD to medical imaging for dementia classification from MRI scans. The image region is divided into $2\times2$ patches, each encoded into a trainable 2-qubit quantum circuit, and the resulting quantum feature map is processed by a lightweight CNN. The student is evaluated on ADNI-1, ADNI-2, and OASIS-2 datasets. Unlike most C2Q studies, distillation reduced the student's accuracy: without distillation, the student achieved 99.78\%, 98.36\%, and 96.33\%, whereas with distillation, the student achieved 95.23\%, 96.11\%, and 94.12\%.\footnote{The abstract of the source reverses the with- and without-KD values; the values reported here are obtained from the results tables of the paper (ADNI-1 and ADNI-2 values without KD are swapped in the conclusion).}

Franco et al. \cite{franco2026foundation} use ECGFounder, a high-capacity model pretrained on electrocardiogram (ECG) recordings and developed for a range of clinical ECG tasks, as the teacher, fine-tuned for binary arrhythmia classification on the MIT-BIH and PTB-XL datasets, and distilled into a quantum student. The quantum student uses a convolutional autoencoder to compress each ECG window into 6 features, which are then processed by a 6-qubit VQC with 36 trainable parameters. For comparison, two classical students, ResNet-1D and CNN-1D, are distilled from the same teacher. The quantum student performs comparably to the classical students, achieving the highest accuracy among the students on MIT-BIH (68.79\% at $T=2$) and PTB-XL (75.03\% at $T=4$), while the teacher still performs better. However, because the study does not include a student trained without distillation, it does not show whether distillation itself improves the quantum student.

C2Q also extends to structured prediction. Roh et al. \cite{roh2024fast} distill a pretrained classical region proposal network (RPN), which proposes bounding boxes around potential objects, into a quantum-based RPN built from 4-qubit $2\times2$ and single-qubit $1\times1$ filters. Evaluated on KITTI, an autonomous-driving object-detection benchmark, the detector with 64 channels achieves $\mathrm{mAP}@0.5$ of 51.2 with distillation, versus 21.1 without KD, but remains below the classical Faster R-CNN baseline (68.4). Two follow-up studies also report gains from distillation. Kim et al. \cite{kim2025quantum} report $\mathrm{mAP}@0.5$ of 41.2 with distillation, versus 12.6 without, but do not compare with a classical detector. Roh et al. \cite{roh2025hybrid} extend the approach to 3D object detection by distilling the teacher's predicted heatmaps, raising average precision at an intersection-over-union (IoU) threshold of 0.5 from 58.22 to 79.35 at moderate difficulty, which is close to the classical baselines (80.37 and 82.89), but a larger gap remains at an IoU of 0.7.

To address the constraint of language tasks where the teacher's capacity exceeds that of the student, Li et al. \cite{li2025quantum} proposed a KD framework in which a fine-tuned LLM transfers its knowledge to a small-scale quantum model through KD. Given the limited capacity of quantum hardware, the NLP task is reformulated as a classification task, thereby enabling an 11-qubit quantum model to learn effectively from an LLM teacher\footnote{The authors describe the model as a quantum-inspired classical algorithm; we classify it as C2Q because its student is a PQC evaluated in simulation and on quantum hardware.} This study notes that the quantum student's capacity is bounded by the qubit count and circuit expressivity, limiting the framework's applicability to complex NLP tasks.

While earlier LLM frameworks focus on distilling output logits into small VQCs, Aizpurua et al. \cite{aizpurua2026quantum} introduce Cayley unitary adapters into a compressed SmolLM2-135M model, trained with distillation from the uncompressed model to recover part of the lost performance. Compression raises perplexity on WikiText from 24.10 to 35.31. In a noiseless simulation, 2-qubit blocks reduce perplexity to 34.68. The adapters are trained classically and executed on quantum hardware only at inference. In a simulation with IBM Heron~r2 noise, perplexity rises to 45.50 with 2-qubit blocks, above the compressed backbone, and 1,189 with 3-qubit blocks.

In another federated learning setting, an end-to-end learning-based system is applied to encode and transmit text over noisy channels \cite{liu2025federated}. This study addresses the challenge of out-of-distribution interference and limited training data, which make generalization difficult. To address this, they proposed a classical teacher-guided hybrid quantum-classical model in which a parameterized quantum circuit serves as the channel encoder-decoder. Results show that adding a quantum channel component slightly improves performance, while robustness under channel interference is attributed to the federated distillation framework itself. In this setting, the clients themselves are quantum models rather than classical models with an embedded quantum component.

Extending the C2Q distillation setting to visual reinforcement learning, Lazaro et al. \cite{lazaro2026staged} propose a staged hybrid framework in which a frozen classical visual teacher provides feature representations to a compact student policy head. The student head, which may be classical or quantum-based, is trained to match the teacher's policy logits using a temperature-scaled KL-divergence objective introduced in Section~\ref{distloss}. CartPole Pixels and Acrobot Pixels serve as pixel-based visual control benchmarks for both student models. Using angle encoding, the quantum-based head achieves returns close to those of the parameter-comparable classical head, although its training time is significantly longer. Generalization across these domains is achieved by adapting the task to fit the student's capacity rather than scaling the student.

Overall, C2Q KD is the most developed direction in the current literature. Most surveyed studies find that guidance from a classical teacher improves student model performance relative to non-distilled baselines \cite{hasan2023bridging,piperno2025classical,ovi2025bridging}. However, the magnitude and reliability of these gains vary across settings, and in some cases distillation does not help at all. Nishan et al. \cite{nishan2026quantummedkd} report significant degradation in 8 of 30 evaluated configurations, Amin et al. \cite{amin2025quantum} obtain lower accuracy with distillation on all three datasets, ED-QTL remains below the strongest transfer heads evaluated by Innan et al. \cite{innan2026towards}, and Boucher et al. \cite{boucher2025n} observe little or negative average benefit for 4-qubit students, with gains only appearing for 8-qubit students. This suggests that the benefits of distillation may depend on the quantum student's capacity. Several studies report gains of less than 1 percentage point, making it difficult to disentangle the effect of distillation from variability due to initialization, optimization, and other architectural choices, including encoding strategies, ansatz expressivity, circuit depth, and classical pre- or post-processing components \cite{hasan2023bridging, ovi2025bridging}. For example, amplitude encoding combined with a fully connected feature-extraction layer yields the best results in Hasan et al. \cite{hasan2023bridging}. This suggests that the effectiveness of the distillation not only depends on the teacher's signal, but also on how the student represents and uses the transferred knowledge. Comparative results also suggest that fully quantum models can obtain larger relative gains from distillation than hybrid models \cite{piperno2025classical}. One possible explanation is that hybrid models already benefit from classical components that support learning, leaving less room for improvement through distillation. More robust ablation studies are therefore needed to isolate the specific impact of different architectural choices and the distillation signal itself.

From a loss perspective, response-based KL divergence loss is the dominant distillation objective across C2Q studies, consistent with the hybrid output outline described in Section~\ref{sec:qkd}, where most frameworks employ a classical readout followed by temperature-scaling \cite{hasan2023bridging, piperno2024quantum, piperno2025classical, 
nishan2026quantummedkd, li2024hybrid, ovi2025bridging, liu2025federated, lazaro2026staged}. However, Li et al. \cite{li2025quantum} incorporate Jensen-Shannon divergence as a complementary divergence measure. Feature-based KD is represented by Barbato et al. \cite{barbato2026quantum}, Roh et al.'s HQCOD \cite{roh2025hybrid}, and Boucher et al. \cite{boucher2025n}, while Roh et al.'s QODE \cite{roh2024fast} and Kim et al. \cite{kim2025quantum} apply a response-based objective to logits. No relation-based loss was identified among the included C2Q studies.

\subsection{Quantum-to-Quantum KD (Q2Q)}
KD in a quantum-to-quantum setting focuses on transferring knowledge between two quantum models. In many Q2Q KD frameworks, the student model is used as a compressed version of the teacher model, using fewer qubits or a shallower circuit. This approach is advantageous because, as the system's scale increases, deploying a large-scale quantum model becomes impractical.

Yan et al. \cite{yan2026distilling} explored this setting by distilling a large quantum model into a smaller one. To reduce the qubit count of the quantum input data, a quantum autoencoder (QAE) is employed to lower the dimensionality of the teacher input before the compressed states are passed to the student. Experiments conducted on both classical and quantum data demonstrated that compressed quantum models achieve improved performance compared to the same undistilled students, with the largest gains observed in topological-phase classification, where accuracy increases by 29.40 percentage points using a 2-qubit circuit with 2 layers, while a 7-qubit circuit with 1 layer beats the teacher, reaching 99.8\% versus 98.3\%. In addition to model compression, this study explores what the authors call self-distillation, which uses identical teacher and student model structures. Rather than compressing the model size, self-distillation focuses on accelerating convergence during training.

Alam et al. \cite{alam2023knowledge} investigated Q2Q KD through approximate synthesis. The proposed framework differs from conventional distillation methods by treating the parameterized quantum circuit (PQC) of a trained quantum neural network (QNN) as the source of knowledge. A student is constructed by approximating the teacher model's unitary transformations. The student is trained to mimic the teacher's unitary transformations, producing a similar circuit with fewer layers and fewer gates. This type of circuit-level imitation can be expressed through a unitary-overlap objective of the form
\begin{equation}
d = 1 - \frac{\left|\operatorname{Tr}(U^\dagger V)\right|}{\dim(U)}
\end{equation}
where $U$ and $V$ denote the unitary matrix representations of the teacher and student circuits, respectively. When the student circuit closely approximates the teacher circuit, $U^\dagger V$ approaches the identity matrix up to a global phase, which does not affect any measurement outcome, and the distance $d$ approaches zero. This formulation differs from standard soft-label KD because the student approximates the quantum operation represented by the teacher circuit, not just the output probabilities, allowing the model to be compressed or adapted to different quantum hardware platforms without full retraining. Averaged across all datasets and two emulators (FakeMelbourne and FakeAlmaden), the four-layer student obtained by approximate synthesis followed by two epochs of supervised fine-tuning achieved 67.7\% mean validation accuracy, compared with 57.3\% for the seven-layer base model.

Extending circuit approximation beyond classification, Piperno et al. \cite{piperno2026knowledge} distill a ten-layer multi-angle QAOA teacher into a single-layer QAOA student, where the student approximates the teacher's outputs rather than its unitary. The study compares three distillation schemes: KL divergence between measurement probability distributions, MSE between per-qubit $\langle Z\rangle$ expectation values, and Huber loss between the same expectation values. Across 16 MaxCut graphs with 4 to 10 nodes, MSE and Huber distillation either maintain or improve performance relative to non-distilled students, while KL-based distillation shows high variance and inconsistent performance. However, it is unclear how much teacher information the MSE and Huber targets carry, since the $\langle Z_i\rangle$ values are expected to be zero by symmetry in the MaxCut setting. Both the teacher and the student are optimized separately for each graph, so the evaluation concerns compression for individual problem instances; transfer to unseen graphs is not assessed.

Krishnamurthy et al. \cite{krishnamurthy2026early}\footnote{
The paper describes the KD framework from a quantum teacher to a quantum student, hence a Q2Q classification. The teacher incorporates CNN-LSTM components together with a PQC, making it a hybrid quantum-classical architecture; the paper title, however, characterizes the framework as quantum-enhanced.} propose a Q2Q KD framework in which a high-capacity hybrid quantum-classical CNN-LSTM teacher supervises a lightweight quantum student model, designed for the early detection of DDoS attacks in distributed cloud environments. The teacher model uses a parameterized quantum circuit with angle encoding to capture complex relationships among traffic features and hidden attack patterns that classical models might overlook. Using the CIC-DDoS2019 dataset, the proposed method achieves an attack recovery ratio of 99\%, a packet loss rate of 20\%, and an attack detection time of approximately 40~ms. However, the study's reported results are not internally consistent, as the comparison and ablation tables list 0.4\% packet loss and 30~ms attack detection time for the same framework. This study extends Q2Q KD beyond circuit compression to data and cloud security settings.

Beyond classification, Zhu et al. \cite{zhu2025qctkd} extend quantum KD to point cloud upsampling, in which a compact student learns from a larger teacher to generate denser 3D point sets while preserving the shape of the represented object.\footnote{The teacher is described inconsistently: Fig.~1 in this study depicts a classical teacher, while Table~V marks it as using quantum convolution. We provisionally classify the work as Q2Q based on Table~V. The distillation loss is unspecified, and no comparison with the same quantum student trained without KD is provided.} The student's feature extractor is a quantum convolutional transformer that encodes features with $R_X$ and $R_Y$ rotation gates and uses data re-uploading to process high-dimensional features with a limited number of qubits. On PU1K with $4\times$ upsampling, the student has a reported model size of 1.7\,M versus 2.4\,M for the teacher and an inference time of 2.9\,ms versus 3.3\,ms. However, its Chamfer distance, which measures the discrepancy between the model's generated and ground-truth point clouds, is higher (0.359 $\times10^{-3}$ versus 0.215 $\times10^{-3}$).

Yu et al. \cite{yu2025knowledge} extend Q2Q KD to a distributed learning setting, addressing the challenges of data and model heterogeneity in federated quantum environments. Instead of sharing model parameters, local quantum models share their predictions on a small public dataset held by the server. This highlights that, in federated Q2Q KD, knowledge transfer occurs through aligning predictions rather than directly sharing model parameters. Gurung and Pokhrel \cite{gurung2025llm} proposed LLM-QFL, in which clients refine a local quantum model towards the aggregated quantum global model, while a fine-tuned LLM teacher guides the quantum clients' optimizer to improve communication efficiency and convergence. Final result accuracies on the benchmark datasets are close to chance.

Unlike the usual setting in which a teacher guides a smaller student, Han et al. \cite{han2026layerwise} use a classifier at every depth of a single quantum model, combining binary cross-entropy at each depth with pairwise KL divergence between the predicted probability distributions, so that each classifier acts as both teacher and student. This includes what the authors call reverse distillation, where shallower classifiers guide deeper ones. Their framework targets federated learning, in which clients run circuits of varying depths based on their hardware capacity. The authors report up to 12.4 pp improvement in testing accuracy over Q-HeteroFL, a heterogeneous federated learning method adapted from classical ML. Comparisons with standalone and vanilla quantum federated learning show smaller gains and occasional underperformance. Hardware experiments on six IBM processors, using 100 test points for Fashion-MNIST pullover/coat classification, reported that the Quorus-Funnel model achieved 48-76\% accuracy, compared with 79\% in ideal simulation.

Similarly, Oh et al. \cite{oh2025quantum} propose quantum infidelity codistillation, in which four quantum students are trained jointly. During joint training, each student provides supervision to its peers. Each student is trained with a supervised loss and a regularization term, $1-|\langle\phi^{j}|\phi^{j'}\rangle|$, that aligns each student's quantum state with those of its peers before measurement. On MNIST, accuracy is achieved similar to teacher-supervised quantum KD (60.76\% versus 60.66\%), although mean accuracy is lower (57.18\% versus 58.60\%).

Unlike C2Q KD, Q2Q approaches serve several purposes beyond model compression. The motivation for this distillation direction is to support circuit compression or approximation, improve convergence, and facilitate knowledge exchange in federated settings, while still preserving a high performance similar to that of the teacher model. Because both models are quantum, this is the only direction in which a student can learn directly from another model's circuit or quantum states rather than from outputs, as in Alam et al. \cite{alam2023knowledge}, who approximate the teacher's unitary, and Oh et al. \cite{oh2025quantum} who align pre-measurement states. However, each study uses a different reference against which to measure its gain. Yan et al. \cite{yan2026distilling}, Piperno et al. \cite{piperno2026knowledge}, and Yu et al. \cite{yu2025knowledge} compare their models against a model trained without distillation, while others compare against the teacher \cite{alam2023knowledge} or other detection methods \cite{krishnamurthy2026early}. As a result, the studies remain fragmented, making a controlled comparison of performance, resource cost, and robustness under realistic quantum device noise models largely absent.

\subsection{Quantum-to-Classical KD (Q2C)}
Quantum-to-classical KD refers to the distillation of knowledge from a quantum or hybrid quantum-classical model to a classical model. In principle, this direction enables the transfer of information learned by a quantum model into a deployable classical model. This is particularly relevant for near-term settings in which quantum resources are available during training, while inference must ultimately be performed classically on conventional hardware.

Despite its potential, Q2C remains the least explored direction in the current literature. Naqvi et al. \cite{naqvi2026structured} present a framework in which a VQC distills its knowledge to lightweight classical students, including a CNN and an MLP. The framework first applied average pooling to each input image, compressing it into a fixed-length vector that can be processed by the VQC. A quantum Fourier transform (QFT)-inspired encoding is then applied to map the compressed features into circuit rotations, providing temperature-scaled soft labels to guide the classical student.

Naqvi et al. \cite{naqvi2026structured} report 43.27\% clean MNIST accuracy for the quantum teacher and 96.50\% for the distilled classical student. The reported robustness benefit concerns performance averaged over a combined Gaussian-noise–rotation grid and does not establish superiority over classical KD under Gaussian noise alone.

Majeed et al. \cite{majeed2026compute} apply Q2C distillation to reinforcement learning, distilling a variational quantum policy into a lightweight classical proxy policy for UAV navigation, using an online scheme; the classical proxy is used in low-uncertainty states, while the quantum policy is still used in high-uncertainty states. The proxy policy minimizes the squared distance between the mean actions of the classical and the quantum policies. However, no ablation isolates the effect of distillation on the proxy.

Therefore, whether a quantum or hybrid-quantum teacher can provide a more informative supervisory signal than that of a suitably matched classical teacher remains an open question due to the limited studies in this direction. Addressing this question requires a controlled comparison with a classical teacher with a comparable number of parameters and level of predictive performance. This suggests that the direction is an open research area requiring further exploration.

\subsection{Quantum-Enhanced Distillation}
\label{subsec:qenhanced}
In contrast to studies that define quantum KD through teacher- or student-architecture, this category is characterized by quantum resources that enhance the distillation process itself when neither the teacher nor the student is quantum. Rather than focusing on the KD direction, these approaches leverage quantum kernels, circuits, or quantum hardware to enhance the representation and transfer of knowledge during distillation.

A prominent work is quantum relational knowledge distillation (QRKD) \cite{liu2025quantum}, in which Liu et al. introduce a framework that maps classical features into a high-dimensional Hilbert space to compute quantum kernel values. This framework combines response-based KD with distance- and angle-based relations to capture complex inter-sample relationships that may be difficult to represent in a conventional classical feature space. Experiments on vision benchmarks (MNIST and CIFAR-10) and language tasks (WikiText-2, Penn Treebank, and IMDB) show that gains depend on the task and loss configurations, while both the teacher and student models remain deployable on classical hardware.

Similarly, Chen et al. \cite{chen2026photonic} adopt a classical-to-classical KD framework, leveraging quantum resources during student training to enhance the model. More specifically, a simulated continuous-variable photonic sampler is used to generate conditioning features for dictionary convolutions, restricting the trainable convolutional parameterization to a low-dimensional family. The reported experiments are run on the ORCA simulator backend in CUDA-Q, with an interface designed to support subsequent hardware execution. Performance varies with task complexity: the compressed student matches the teacher on MNIST (99.09\% versus 99.07\%) and FashionMNIST (92.42\% versus 91.86\%), but falls below the teacher on CIFAR-10 (79.72\% versus 86.99\%).

Extending this to language-model compression, Ni et al. \cite{ni2026quloc} use photon-detection probabilities to generate channel-wise scaling coefficients for low-rank weight factors, trained through distillation from a larger teacher. These coefficients are absorbed into the weights after training, leaving inference classical. On Qwen3.5-4B, average downstream accuracy reaches 54.03\% versus 48.54\% for a matched ungated baseline.

While the above works use quantum resources to enrich knowledge representation, Li et al. \cite{li2026quantumgated} use them to weight the transfer between the classical teacher and student. In their class-incremental learning framework, a frozen pretrained model learns each new task through a small adapter, which is distilled from previous adapters. A parameterized circuit encodes the current sample, and each past task is represented by a normalized summary of its adapter's principal directions, and their softmax-normalized fidelity sets how much that task adapter contributes to the distillation loss. The quantum gate also remains active at inference, where its sample-dependent relevance weights combine or select task-specific adapters. The method outperforms twelve continual learning baselines on five benchmarks; against classical routing alternatives, quantum gating reaches 81.79\% versus 80.08--80.32\% average incremental accuracy on ImageNet-R and 93.54\% versus 92.57--92.72\% on CUB-200 with higher per-image inference latency.

This category is conceptually different from the teacher-student configurations discussed in the previous sections. Rather than defining the quantum component as either the teacher or the student, these approaches use quantum resources to enhance the distillation of classical models. However, whether this direction offers a consistent advantage over purely classical KD methods remains an open question, especially when additional computational and hardware costs are taken into account. 

\subsection{Quantum-Inspired Distillation}
\label{subsec:qinspired}
Quantum-inspired KD is a conceptually distinct category that denotes the teacher-student transfer performed entirely on classical hardware, in which the teacher, the student, or the distillation mechanism employs a quantum-derived mathematical construction, without evaluating any quantum state, circuit, or kernel at any stage. Unlike teacher-student directions in which at least one model simulates or executes quantum gates, quantum-inspired KD employs classical formulations inspired by quantum information theory, such as activations modeled after qubit rotations or tensor networks originating in quantum many-body physics. Within this category, KD is used to train these architectures, while both training and inference remain fully classical. We include only studies that explicitly derive their methods from concepts in quantum physics or quantum information theory. Examples include constructions based on matrix product operators from quantum many-body physics and methods that draw on analogies with quantum rotations or interference. We exclude studies that present tensor decompositions solely as numerical techniques.

An example is the slice-wise feature distillation framework proposed by Hamreras et al. \cite{hamreras2026fast}, where a pretrained network is split into slices, convolutional layers are factorized with Tucker decomposition and MLP blocks with matrix product operators, and each slice is optimized to reproduce the intermediate features of its dense pre-trained counterpart. On ResNet-34, this scheme outperforms the fine-tuning of the tensorized model without distillation (94.7\% versus 89.47\% on CIFAR-10 at a compression rate of 0.5) while requiring less than half the optimization time. On the GPT-2 XL model, this does not carry over; slice-wise distillation is better on the LAMBADA dataset, while end-to-end fine-tuning is better on WikiText and C4.

Quantum-inspired work extends tensor-network distillation to other architectures and training objectives. Murua et al. \cite{murua2020tensor} factorize the weight matrices of an LSTM cell into MPS and MPO formats, where the weights or activations of the uncompressed model serve as supervisory signals for the tensorized cell. The study’s results show that distilling the weights closely matches the uncompressed model on Penn Treebank at low compression (79.95 versus 79.91 perplexity), whereas distilling the activations lowers perplexity below it, as observed on the larger Google One Billion Words dataset (65.90 versus 67.48 perplexity).
A related role appears in Mittal et al. \cite{mittal2026qyolo}, who replace two parameter-heavy backbone modules of YOLOv8 with a channel-mixing block motivated by single-qubit rotations and distill from the uncompressed detector teacher model to recover 0.4 percentage points of $\mathrm{mAP}@0.5$ lost to compression.

Two studies illustrate distinct uses of quantum-inspired methods in distillation. Xu and Muhammad \cite{xu2026quantum}, whose model augments classical attention with quantum-driven self-attention motivated by wave interference, distill for robustness rather than compression, transferring conversational context from a teacher that fuses text, audio, and video to its unimodal branches acting as students, while Zhan et al. \cite{zhan2024over} invert factorization’s usual role, using MPO to over-parameterize a small student during training and merging the tensors afterward so that its inference parameter count is unchanged.

\section{Applications}
\label{sec:application}
This section reviews emerging applications of quantum KD to domain-specific problems beyond standard machine learning benchmarks.
\textbf{Large Language Models.} 
Applying quantum KD to natural language entails dealing with a mismatch of several orders of magnitude between the teacher's representational capacity and the student's. Li et al. \cite{li2025quantum} addressed this by converting the open-ended language task into a closed-set classification with a small label set. This approach allows for an 11-qubit student, but it limits the approach to a classification task. Gurung and Pokhrel \cite{gurung2025llm} address a different constraint in federated settings, where communication is a central concern. Their framework uses a fine-tuned language model teacher to regulate the optimization of quantum clients, aiming to make communication more efficient and improve learning. Overall, these studies showed that current quantum KD approaches for language processing rely on substantially constrained task formulations or distributed learning setups.

Aizpurua et al. \cite{aizpurua2026quantum} extend this to generative language modeling, using distillation to train quantum adapters that recover part of the performance loss incurred during the compression of SmolLM2, with inference executed on IBM hardware.

\textbf{Wireless Communication.}
Semantic communication introduces constraints not present in static classification tasks. The system must remain reliable under novel conditions not observed during training, a challenge that often arises when only limited data are available at individual clients \cite{liu2025federated}. In this context, the quantum component is integrated into the transmission pipeline as a channel encoder-decoder. Therefore, evaluation should focus on robustness to channel interference rather than accuracy on a predetermined test set. Additionally, the field prioritizes compact circuits because the quantum component must operate within the communication system's resource and latency constraints.

\textbf{Cybersecurity.}
Krishnamurthy et al. \cite{krishnamurthy2026early} apply quantum KD to early DDoS detection using the CIC-DDoS2019 dataset. This domain also changes what counts as successful evaluation: reported quantities include attack recovery ratio, packet loss rate, and detection time, rather than classification accuracy alone. Such metrics are not directly comparable with the results reported by most other surveyed studies in Section~\ref{sec:benchmarking}.

\textbf{Healthcare.}
In healthcare, quantum KD has been applied to pediatric pneumonia detection from chest radiographs. Nishan et al. \cite{nishan2026quantummedkd} introduce QuantumMedKD, a framework that integrates hybrid quantum-classical models with knowledge distillation for medical image analysis to detect pediatric pneumonia. The study is motivated by resource-constrained clinical settings and limited data, where compact models and additional teacher-guided supervision may be useful, even when the principal goal is not maximum predictive accuracy. 

Quantum KD has also been applied to other clinical data. Amin et al. \cite{amin2025quantum} distill a CNN teacher into a hybrid quantum student for dementia classification from MRI scans, although the student achieves lower accuracy across all three datasets than without distillation. Franco et al. \cite{franco2026foundation} distill an ECG foundation model into compact students for arrhythmia detection, showing that a quantum student with only 36 trainable circuit parameters can match classical students distilled from the same teacher, although all students yield lower accuracy than the teacher.

\textbf{Image Classification and Visual Reinforcement Learning.} 
Quantum KD has also been evaluated in image classification and visual reinforcement learning. Barbato et al. \cite{barbato2026quantum} apply feature-based distillation to compress a CNN teacher into a hybrid student for binary and multi-class MNIST and Fashion-MNIST classification tasks. Instead of matching the teacher’s final output distribution, the student learns by aligning intermediate feature representations, a different approach from response-based KD. Lazaro et al. \cite{lazaro2026staged}, on the other hand, extend quantum KD to visual reinforcement learning using CartPole pixels and Acrobot pixels to evaluate compact, classical, and quantum-based policy heads. This study reports that quantum policy heads on CartPole achieve returns comparable to their classical counterparts but with longer simulated training (66.8 vs. 6.3 minutes).

\textbf{Object Detection and 3D Vision.}
Quantum KD has also been applied to object detection and 3D vision. Roh et al. \cite{roh2024fast} and Kim et al. \cite{kim2025quantum} distill a classical region proposal network into a quantum counterpart for object detection in autonomous driving, and Roh et al. \cite{roh2025hybrid} extend this study to 3D detection, while Zhu et al. \cite{zhu2025qctkd} apply KD to point cloud upsampling. For object detection, evaluation is performed using mean average precision ($\mathrm{mAP}@0.5$), while point cloud upsampling is evaluated using geometric error metrics between the generated and ground-truth point sets, such as Chamfer distance.

\begin{table*}[!h]
\centering
\caption{Multi-dimensional cross-reference of quantum KD frameworks}
\label{tab:crossref}
\small
\begin{tabular}{p{0.08\textwidth} 
>{\raggedright\arraybackslash}p{0.09\textwidth} 
>{\raggedright\arraybackslash}p{0.16\textwidth}
>{\raggedright\arraybackslash}p{0.17\textwidth}
>{\raggedright\arraybackslash}p{0.16\textwidth}
>{\raggedright\arraybackslash}p{0.14\textwidth}}
\hline\\
Paper & KD Direction & Quantum Model Type & Distillation Target & Distillation Mechanism & Main Goal  \\
\hline\\

\cite{hasan2023bridging}
& C2Q & Hybrid & Softened class probabilities & Response-based & Performance \\[2pt]

\cite{piperno2025classical}
& C2Q & Fully quantum / Hybrid & Class probabilities / Logits & Response-based & Performance \\

\cite{piperno2024quantum}
& C2Q & Hybrid & Class probabilities & Response-based & Performance \\

\cite{innan2026towards} & C2Q & Hybrid (frozen CNN + PQC head) & Logits (teacher ensemble) & Response-based & Performance \\

\cite{li2024hybrid}
& C2Q & Continuous-variable & Class probabilities & Response-based & Performance \\

\cite{ovi2025bridging}
& C2Q & Hybrid & Softened class probabilities & Response-based & Performance \\

\cite{nishan2026quantummedkd}
& C2Q & Hybrid & Softened class probabilities & Response-based & Performance + compression \\

\cite{amin2025quantum} & C2Q & Hybrid (QCNN + CNN) & Logits & Response-based & Performance \\

\cite{franco2026foundation} & C2Q & Hybrid (autoencoder + VQC) & Softened class probabilities & Response-based & Compression \\

\cite{li2025quantum}
& C2Q & Hybrid & Class probabilities & Response-based & Performance  \\

\cite{aizpurua2026quantum} & C2Q & Hybrid (LLM + unitary adapters) & Softened class probabilities & Response-based & Compression (recovery) \\

\cite{liu2025federated}
& C2Q & Hybrid & Intermediate and output distributions & Response- and feature-based & Robustness \\

\cite{barbato2026quantum} & C2Q & Hybrid & Intermediate features & Feature-based & Performance + compression \\

\cite{lazaro2026staged} & C2Q & Hybrid & Softened policy logits & Response-based & Compression \\

\cite{roh2024fast} & C2Q & Hybrid (quantum RPN in Faster R-CNN) & RPN logits & Response-based & Deployment efficiency \\

\cite{kim2025quantum} & C2Q & Hybrid (quantum RPN in Faster R-CNN) & RPN logits & Response-based & Deployment efficiency \\

\cite{roh2025hybrid} & C2Q & Hybrid & Predicted heatmaps & Feature-based & Performance \\

\cite{boucher2025n} & C2Q & Hybrid (ViT with PQC attention) & Intermediate features & Feature-based & Compression (parameter reduction) \\

\cite{yan2026distilling}
& Q2Q & Fully quantum & Softened class probabilities & Response-based & Compression (qubit and depth reduction) \\

\cite{alam2023knowledge}
& Q2Q & Hybrid & Teacher unitary & Circuit-level & Compression \\

\cite{gurung2025llm}
& Q2Q & Hybrid & Global model outputs & Response-based & Communication efficiency \\

\cite{yu2025knowledge}
& Q2Q & Fully quantum & Measurement probabilities & Response-based & Communication efficiency \\

\cite{han2026layerwise} & Q2Q & Fully quantum (multi-depth PQC) & Logits (depth classifiers) & Response-based & Performance \\

\cite{piperno2026knowledge} & Q2Q & Fully quantum (multi-angle QAOA) & Measurement distribution / Pauli-$Z$ expectations & Response-based (KL, MSE, Huber) & Compression (parameter reduction) \\

\cite{krishnamurthy2026early} & Q2Q & Hybrid & Soft labels & Response-based & Performance \\

\cite{zhu2025qctkd} & Q2Q & Hybrid & Refined point clouds & Response-based (loss unspecified) & Compression \\

\cite{oh2025quantum} & Q2Q & Fully quantum (4-qubit PQCs) & Pre-measurement peer states & State-level (infidelity) & Performance (teacher-free) \\

\hline
\end{tabular}
\end{table*}

\begin{table*}[t]
\centering
\addtocounter{table}{-1}
\caption{Multi-dimensional cross-reference of quantum KD frameworks (continued)}
\small
\begin{tabular}{p{0.08\textwidth} 
>{\raggedright\arraybackslash}p{0.09\textwidth} 
>{\raggedright\arraybackslash}p{0.16\textwidth}
>{\raggedright\arraybackslash}p{0.17\textwidth}
>{\raggedright\arraybackslash}p{0.16\textwidth}
>{\raggedright\arraybackslash}p{0.14\textwidth}}
\hline\\
Paper & KD Direction & Quantum Model Type & Distillation Target & Distillation Mechanism & Main Goal  \\
\hline\\

\cite{naqvi2026structured}
& Q2C & Hybrid (VQC + linear readout) & Temperature-scaled soft labels & Response-based & Robustness + Deployment efficiency \\

\cite{majeed2026compute} & Q2C & Fully quantum (VQC policy teacher) & Mean actions & Response-based & Deployment efficiency \\

\cite{liu2025quantum}
& Q-Enhanced & Quantum module (kernel) & Soft labels and pairwise relations & Response- and relation-based & Performance \\

\cite{chen2026photonic}
& Q-Enhanced & Quantum module (Photonic CV sampler) & Softened class probabilities & Response-based & Compression \\

\cite{ni2026quloc} & Q-Enhanced & Quantum module (Photonic gating) & Intermediate features / Top-$K$ token probabilities & Feature- and response-based & Compression \\

\cite{li2026quantumgated} & Q-Enhanced & Quantum module (PQC gating) & Logits (previous adapters) & Response-based & Performance \\
\hline
\end{tabular}
\begin{tablenotes}
  \footnotesize
  \item \textit{Quantum module}: a quantum kernel, sampler, or gate used within a classical distillation pipeline.
  \end{tablenotes}
\end{table*}

\section{Benchmarking}
\label{sec:benchmarking}
Due to the nature of quantum KD and its sensitivity to several factors, it is difficult to establish a fair benchmark for comparison. These factors include: (a) datasets, tasks, and evaluation scales differ across works, (b) qubit counts and circuit architectures vary, and (c) evaluation metrics are non-standardized. In addition, results from ideal simulations cannot be directly compared with those from real quantum hardware or noisy simulators because hardware-specific effects, such as gate errors, decoherence, readout errors, and finite-shot noise, can affect the distillation process. Only six of the 33 surveyed studies are evaluated on physical quantum hardware (Table~\ref{tab:qkd_directions}), while most report on ideal or noise-modeled simulation. Therefore, reported gains should be read according to the claim they support: distillation benefit is shown when the same quantum student is trained with and without KD, a quantum contribution is shown when a distilled quantum student is compared against a parameter-matched classical student distilled from the same teacher, or a compression or deployment benefit is shown when the performance retained relative to the teacher is reported with the resources saved such as parameters, circuit depth or inference time.

Furthermore, another source of incomparability is the baseline choice: some works compare against a non-distilled quantum model of identical architecture, others compare against a classical model of comparable size, and others compare against the teacher itself. Of the 33 surveyed studies, 18 compare against a classical baseline other than the teacher (Table~\ref{tab:qkd_directions}). A gain over an otherwise identical quantum student trained without distillation provides evidence for the benefit of the distillation objective under the evaluated protocol. A gain over a parameter-matched classical student establishes a comparison between the tested pipelines; attributing that gain specifically to the quantum component requires controlled ablations and comparable tuning and resource budgets. The effect of execution environment is also evident within individual studies: Li et al. \cite{li2025quantum} report 86.70\% in simulation for thematic analysis and up to 73.75\% on hardware, and Barbato et al. \cite{barbato2026quantum} report 99.04\% in simulation and 80.50\% for a transpiled and retrained circuit. The authors find that a hybrid student outperforms its parameter-matched classical counterpart in noiseless simulation in five of six binary tasks but only two of four multi-class tasks, whereas Lazaro et al. \cite{lazaro2026staged} observe that a quantum head either matches or underperforms its parameter-matched classical counterpart, requiring substantially more training time. Table~\ref{tab:benchmarking} summarizes these differences across the surveyed studies.

\begin{table*}[ht]
\centering
\caption{Benchmarking of surveyed quantum knowledge distillation works}
\label{tab:benchmarking}
\scriptsize
\renewcommand{\arraystretch}{1.2}
\begin{tabularx}{\textwidth}{p{0.6cm} 
>{\raggedright\arraybackslash}p{1.3cm} 
>{\raggedright\arraybackslash}p{2.6cm} 
>{\raggedright\arraybackslash}X 
>{\raggedright\arraybackslash}X 
>{\raggedright\arraybackslash}p{1.0cm} 
>{\raggedright\arraybackslash}p{1.3cm}}
\toprule
\textbf{Paper} & \textbf{KD Dir.} & \textbf{Dataset} & \textbf{Reported Result} & \textbf{Reported Gain} & \textbf{Qubits} & \textbf{Hardware} \\
\midrule

\cite{hasan2023bridging}$^{\ddagger}$ & C2Q & MNIST / FashionMNIST / CIFAR-10
& 91.27\% accuracy (8q) and 85.18\% (4q) (MNIST,  AlexNet teacher)
& +0.87 pp (8q) to +7.49 pp (4q) vs.\ non-distilled QNN (AlexNet teacher)
& 4, 8 & Simulation \\

\cite{piperno2024quantum} & C2Q & XOR (64-bit, 3-class)
& $\approx$0.47--0.85 F1 (approximation read from figure)
& up to +0.27 F1 vs.\ non-distilled baseline (Universal+H)
& 6 & Simulation \\

\cite{innan2026towards}$^{\ddagger}$ & C2Q & Fashion-MNIST / Hymenoptera (Ants vs.\ Bees)
& 85.45\% (4q), 85.88\% (6q) (Fashion-MNIST); 71.24\% (4q), 74.51\% (6q) (Ants vs.\ Bees)
& Below other QTL heads at 6 qubits (Fashion-MNIST) and at both qubit counts (Ants vs.\ Bees)
& 4, 6 & Simulation \\

\cite{piperno2025classical} & C2Q & MNIST (digits 0--6)
& 0.93--0.95 F1 (hybrid, 8--10-layers), 0.63 F1 (Fully quantum, 4 layers)
& +0.13 F1 vs. non-distilled (4-layer fully quantum; 0.63 vs.\ 0.50)
& not reported & Simulation \\

\cite{li2024hybrid} & C2Q & MNIST / FashionMNIST
& No value - curves only
& reported higher accuracy and faster convergence vs.\ no teacher (curves only)
& 2--5$^{\dagger}$ & Simulation \\

\cite{ovi2025bridging} & C2Q & MNIST (6-class / 8-class)
& 97.78\% accuracy (3q + feature interaction, 6-class)
& vs.\ 96.56\% non-distilled and 97.44\% distilled without feature interaction (3q, 6-class)
& 3, 4 & Simulation \\

\cite{nishan2026quantummedkd}$^{\ddagger}$ & C2Q & Chest X-rays (5{,}863 images) & 84.00\% accuracy (6q, EfficientNet-B0 teacher)
& vs.\ 68.7\% VQC without KD; 71.8\% classical KD student (parameter-matched)
& 3--8 & Simulation \\

\cite{amin2025quantum} & C2Q & ADNI-1 / ADNI-2 / OASIS-2 (MRI)
& 95.23\% (ADNI-1), 96.11\% (ADNI-2), 94.12\% (OASIS-2) (with KD)
&  $-4.55$, $-2.25$, $-2.21$ pp vs. no KD (99.78\%, 98.36\%, 96.33\%)  & 2 & Simulation \\

\cite{franco2026foundation} & C2Q & MIT-BIH / PTB-XL (ECG)
& 68.79\% (MIT-BIH, $T=2$); 75.03\% (PTB-XL, $T=4$)
& Comparable to classical students; teacher 80.00\% / 80.63\%
& 6 & Simulation \\

\cite{li2025quantum} & C2Q & Emotion / Hiding / Thematic (NLP)
& 86.70\% accuracy (thematic analysis, LLaMA2-7B teacher); up to 73.75\% (hardware, sample subset)
& Best on thematic analysis (86.70\% vs.\ 86.21\% for MiniLLM); second to MiniLLM on emotion and hiding detection
& 11 & Simulation + Real (Baihua) \\

\cite{aizpurua2026quantum} & C2Q & WikiText, LAMBADA, BoolQ, HellaSwag, GSM8K, TriviaQA, MMLU & Perplexity 34.68 (2-qubit blocks); 31.25 (8-qubit blocks), noiseless simulation & vs.\ 35.31 compressed model; teacher 24.10; 45.50 with simulated Heron~r2 noise (2-qubit) & 2--10 & Simulation + Real (IBM Heron r2) \\

\cite{liu2025federated} & C2Q & Europarl corpus
& No values - BLEU and sentence similarity curves only
&  reported robust under channel interference vs.\ FL-only baselines; quantum channel coder slightly above classical (curves only)
& 2 & Simulation \\

\cite{barbato2026quantum} & C2Q & MNIST / Fashion-MNIST (6 binary tasks, 4 multi-class)
& 99.04\% (0-versus-9 task, simulation), 80.50\% (retrained transpiled circuit, hardware) & Outperforms parameter-matched classical student on 5 of 6 binary tasks, and 2 of 4 multi-class tasks
& 4, 5 & Simulation + Real (IQM Garnet) \\

\cite{lazaro2026staged}$^{\ddagger}$ & C2Q & CartPole / Acrobot Pixels & 487.2 selected return (CartPole); $-$75.8 (Acrobot)
& Comparable to parameter-matched classical head, with longer simulated training (66.8 vs.\ 6.3 min on CartPole; 53.7 vs.\ 1.3 min on Acrobot Pixels)
& 8, 9 & Simulation \\

\cite{roh2024fast} & C2Q & KITTI (object detection) & 51.2 $\mathrm{mAP}@0.5$ (64 channels, $\gamma=0.3$) & vs.\ 21.1 without KD; below classical Faster R-CNN (68.4) & 1, 4 (per filter) & Simulation \\

\cite{kim2025quantum} & C2Q & KITTI & 41.2 $\mathrm{mAP}@0.5$ (64 channels, $\gamma=0.7$) & vs. 12.6 without KD & 1, 4 (per filter) & Simulation \\

\cite{roh2025hybrid} & C2Q & KITTI & 79.35 3D $\mathrm{AP}@0.5$ (moderate) & vs. 58.22 without KD; classical ResNet18 80.37; Faster R-CNN 82.89 & not reported & Simulation \\

\cite{boucher2025n} & C2Q & 8 biomedical datasets (MedMNIST, OASIS)
& 0.937 accuracy (BloodMNIST, 8-qubit KD student)
& +12.1 pp accuracy against same model with no KD; little or negative average benefit in 4-qubit settings
& 4, 8 & Simulation \\

\hline
\end{tabularx}
\end{table*}

\begin{table*}[ht]
\centering
\addtocounter{table}{-1}
\caption{Benchmarking of surveyed quantum knowledge distillation works (continued)}
\scriptsize
\renewcommand{\arraystretch}{1.2}
\begin{tabularx}{\textwidth}{p{0.6cm} 
>{\raggedright\arraybackslash}p{1.3cm} 
>{\raggedright\arraybackslash}p{2.6cm} 
>{\raggedright\arraybackslash}X 
>{\raggedright\arraybackslash}X 
>{\raggedright\arraybackslash}p{1.0cm} 
>{\raggedright\arraybackslash}p{1.3cm}}
\toprule
\textbf{Paper} & \textbf{KD Dir.} & \textbf{Dataset} & \textbf{Reported Result} & \textbf{Reported Gain} & \textbf{Qubits} & \textbf{Hardware} \\
\midrule

\cite{yan2026distilling} & Q2Q & MNIST / topological-phase data
& 81.7\% (2-qubit student) and 99.8\% (1-layer student), topological-phase; teacher 98.3\%
& +29.4 pp (qubit reduction) and +13.8 pp (depth reduction) vs. non-distilled
& 2, 7, 10, 15 & Simulation \\

\cite{alam2023knowledge} & Q2Q & MNIST / FashionMNIST (3-class) & 67.7\% average validation accuracy (4-layer student, fine-tuned)
& +10.4 points vs.\ 7-layer base model, under noise
& 4 & Emulator (FakeMelbourne and FakeAlmaden) \\

\cite{gurung2025llm} & Q2Q & DemoHumanOrWorm / TweetEval-Sentiment
& $\approx$0.56 accuracy (simulation); $\approx$0.47 (hardware; 3 communication rounds)
& No reported accuracy gain
& 4 & Emulator + Real (IBM Brisbane) \\

\cite{krishnamurthy2026early} & Q2Q & CIC-DDoS2019
& 20\% packet loss rate; 99\% attack recovery ratio (from text)
& vs.\ 50--81\% packet loss for classical baselines
& 16 & Simulation \\

\cite{piperno2026knowledge} & Q2Q & MaxCut (4 graph families, 4--10 nodes)
& Approximation ratio (figures only)
& Expectation-value KD never below non-distilled QAOA; count-based KD inconsistent
& 4--10 & Simulation \\

\cite{yu2025knowledge} & Q2Q & MNIST (10-class, heterogeneous)
& 95.02\% accuracy (global model)
& +5.22 pp vs.\ public-data hard-label baseline; +71.98 pp vs.\ average local-client accuracy & 8 & Simulation \\

\cite{han2026layerwise}$^{\ddagger}$ & Q2Q & MNIST / Fashion-MNIST (binary) & Up to 99.4\% (Quorus-Layerwise, Trouser/Boot, 6-layer clients, simulation); 76\% on IBM Torino (Quorus-Funnel, Pullover/Coat, 100 test points) & +12.4 pp over Q-HeteroFL, smaller or negative gains vs.\ standalone/vanilla QFL & 10 & Simulation + Real (IBM) \\

\cite{zhu2025qctkd} & Q2Q & PU1K / PU-GAN (4$\times$ upsampling)
& CD $0.359\times10^{-3}$ (student, PU1K)
& vs.\ $0.215\times10^{-3}$ for teacher; no non-distilled quantum control
& not reported & not reported \\

\cite{oh2025quantum}$^{\ddagger}$ & Q2Q & MNIST
& 60.76\% top-1 accuracy
& vs.\ 60.66\% for teacher-supervised QKD; mean 57.18\% vs.\ 58.60\%
& 4 & Simulation \\

\cite{naqvi2026structured} & Q2C & MNIST
& 96.50\% student accuracy (teacher: 43.27\%)
& vs.\ 98.73\% for classical KD, reported retention: 0.774 vs. 0.732; approximately +0.042 calculated from source's Table 9
& 8, 9, 10 & Simulation \\

\cite{majeed2026compute} & Q2C & UAV navigation & Moving average return $-300$ to $-340$ & No distillation ablation & not reported & Simulation \\

\cite{liu2025quantum}$^{\ddagger}$ & Q-Enhanced & MNIST / CIFAR-10 / WikiText-2
& 95.38\% accuracy (MNIST); reported perplexity 2.3409 (WikiText-2, FQK) & RKD: 95.07\%; no-KD student: 94.91\% (MNIST); RKD perplexity: 2.0183 (WikiText-2)
& 4 & Simulation \\

\cite{chen2026photonic}$^{\ddagger}$ & Q-Enhanced & MNIST / FashionMNIST / CIFAR-10 & 99.09\% / 92.42\% / 79.72\% validation accuracy & vs.\ teacher 99.07\% / 91.86\% / 86.99\% & 16 modes & Simulation \\

\cite{ni2026quloc} & Q-Enhanced & WikiText-2 / C4 (perplexity); ARC-Easy, PIQA, OpenBookQA, WinoGrande, HellaSwag, MathQA (accuracy) & 54.03\% average accuracy
& vs.\ 48.54\% for matched ungated baseline; 53.23\% using measured photonic responses
& 16 modes & Simulation + Photonic quantum hardware (4-modes) \\

\cite{li2026quantumgated} & Q-Enhanced & CIFAR-100 / CUB-200 / ImageNet-A / ImageNet-R / VTAB & 94.08\% / 93.54\% / 65.77\% / 81.79\% / 95.81\% (avg.\ acc.) & +0.03--2.77 over strongest baselines & 3--15 (sensitivity sweep) & Simulation \\
\bottomrule
\end{tabularx}
\begin{tablenotes}
  \footnotesize
  \item \textit{Reported Result}: best configuration reported by the authors unless stated otherwise.
  \item ``Modes'' refers to continuous-variable photonic quantum modes rather than discrete qubits.
  \item ``Emulator'' denotes a noise-model simulation of a named device, not execution on physical quantum hardware.
  \item $^{\ddagger}$Simulation results in this row are means over repeated runs or data samples; the source also reports their variability.
  \item $^{\dagger}$Li et al.\ report swept circuit size in qubits, although the model is continuous variable and measurement step is in qmodes.
\end{tablenotes}
\end{table*}

Although training costs vary substantially, few surveyed studies document them. Students trained in simulation can benefit from efficient differentiation of the simulated state, whereas hardware-based training can use methods such as the parameter-shift rule, which requires two circuit evaluations per parameter to compute each gradient, under the assumption in Eq.~\ref{eq:paramshift}. Barbato et al. \cite{barbato2026quantum} found that training with the parameter-shift rule was approximately 8500 times slower than training the classical baseline in simulation. Similarly, Lazaro et al. \cite{lazaro2026staged} observed that the quantum policy head on CartPole takes approximately 10 times longer to train than the parameter-matched classical head, and approximately 41 times longer on Acrobot. Other studies present accuracy results without specifying the corresponding training costs, making it difficult to assess whether quantum KD is practical beyond the settings they evaluated.

Several C2Q studies show gains of less than 1 percentage point in some configurations. This might be hard to distinguish from differences caused by parameter setup, finite-shot sampling, or training stochasticity unless the studies report repeated runs and uncertainty estimates. Therefore, controlled baseline differences, confidence intervals, and statistical tests should be presented consistently; absolute accuracy alone is inadequate to show the practical impact attributed to KD.

\section{Open Problems}
\label{sec:challenges}
Despite recent advances, quantum KD remains an emerging field. The following challenges continue to limit its \mbox{development:}

\noindent\textit{Output Mismatch.}
A fundamental challenge in quantum KD is the heterogeneity in outputs of quantum and classical models. Classical models output logits that can be temperature-scaled and then passed through the softmax; quantum models can produce several output types, each requiring a different procedure before applying the distillation loss. As discussed in Section~\ref{q-softtargets}, quantum models can produce measurement probabilities that can be treated directly as probability distributions, or expectation values that sometimes require an additional mapping or post-processing step before the classification loss is computed. Hybrid models partially resolve this mismatch by introducing a classical readout layer but also add additional trainable parameters, which might obscure the contribution of the quantum component.

\noindent\textit{Hardware Constraints and Noise Sensitivity.} 
A practical deployment of quantum KD is constrained by the current NISQ hardware \cite{bharti2022noisy}. This includes restricted qubit count, limited circuit depth, and decoherence. These constraints directly affect the quantum model's capacity and outputs during training, whether for the teacher or the student. Furthermore, the number of measurement shots per circuit execution introduces noise to the distillation signals. Among the surveyed studies, only a subset explicitly evaluate their framework under noise: Alam et al. \cite{alam2023knowledge} tested approximate synthesis under emulated device noise, Chen et al. \cite{chen2026photonic} conducted perturbation studies including finite-shot noise, feature corruption, and parameter drift, Li et al. \cite{li2025quantum} deployed their student on the Baihua superconducting processor to report performance under decoherence and finite sampling; and Gurung and Pokhrel \cite{gurung2025llm} validated their LLM-QFL framework on real IBM quantum hardware. Barbato et al. \cite{barbato2026quantum} ran a transpiled, retrained version of their student on the IQM Garnet processor; Aizpurua et al. \cite{aizpurua2026quantum} executed 2-qubit adapters on IBM hardware; and Han et al. \cite{han2026layerwise} evaluated their framework on six IBM processors. Ni et al. \cite{ni2026quloc} use four-mode photonic circuits to measure photon detection probabilities from which the gating coefficients are constructed. Yan et al. propose integrating quantum KD with error-correction protocols, offering a promising direction for improving the distilled model on realistic hardware \cite{yan2026distilling}. Furthermore, the global approximate-synthesis implementation studied in \cite{alam2023knowledge} exhibits increasing approximation error in its tested two-to-six-qubit settings. Explicit unitary representations also require exponential classical storage. Block-wise synthesis offers a possible extension to larger circuits, although scalable distillation performance remains to be demonstrated.

\noindent\textit{Evaluation and Attribution Difficulty.} 
A recurring issue across the surveyed literature is the difficulty of attributing performance gains to the distillation process. Several factors beyond distillation may affect performance, such as data encoding choices, ansatz design or entangling patterns, classical pre- or post-processing choices, optimization settings, or trainable parameter count. This makes it difficult to determine whether the improved performance is due to the distillation itself or to architectural choices in the quantum or hybrid models. The supporting evidence remains limited; a small subset of the surveyed literature evaluates their framework on quantum hardware or a classical counterpart under matched conditions. This inconsistency makes it difficult to derive a unified conclusion regarding quantum advantage in the KD context. Barbato et al. \cite{barbato2026quantum} find that a hybrid quantum-classical student, under noiseless simulation, outperforms a classical student of the same size, whereas Lazaro et al. \cite{lazaro2026staged} observe that classical heads with matching parameters either match or exceed quantum heads while requiring substantially less training time. These contrasting outcomes imply that conclusions depend on the task type, the noise conditions, and the inclusion of computational overhead. Hasan et al. \cite{hasan2023bridging} partially address this by conducting ablation studies that evaluate encoding strategies across multiple datasets. Roh et al. \cite{roh2024fast} provide a direct ablation by holding the student architecture fixed, although the distillation coefficient also rescales the task loss term. However, systematic benchmarking with controlled ablation across different distillation components remains limited; therefore, a standardized evaluation protocol for quantum KD would improve comparative studies in the future.

\noindent\textit{Limited Exploration of Understudied Directions.} 
Current quantum KD studies are heavily focused on C2Q distillation, while other directions remain underexplored. In particular, Q2C distillation, despite its potential to leverage quantum principles (e.g., quantum kernel structure, entanglement-based representations) that can be distilled into deployable classical models, has received limited attention. Similarly, the potential of quantum-enhanced distillation remains an emerging area with limited empirical evidence.

\noindent\textit{Distillation Mechanism Homogeneity.} 
The surveyed literature can be organized according to five distillation targets: response-level outputs, intermediate feature representations, relational structures, unitary transformations, and pre-measurement quantum states. Although response-based distillation remains the dominant approach, alternative distillation targets are represented in the surveyed literature. Feature-based distillation is represented by Barbato et al. \cite{barbato2026quantum} and Boucher et al. \cite{boucher2025n}, circuit-level unitary transfer by Alam et al. \cite{alam2023knowledge}, relation-based quantum-enhanced distillation by Liu et al. \cite{liu2025quantum}, and state-level transfer by Oh et al. \cite{oh2025quantum}. Thus, non-response distillation remains less extensively explored in the surveyed corpus. This poses a mechanism diversity gap that limits current understanding of which distillation target is most effective across different teacher-student configurations.

\noindent\textit{Task Diversity.}
The majority of KD in quantum settings has been evaluated on classification tasks, typically binary or small multi-class classification problems, while non-classification tasks remain sparsely represented. The exceptions include reinforcement learning, where Lazaro et al. \cite{lazaro2026staged}, and Majeed et al. \cite{majeed2026compute} apply policy distillation to pixel-based control tasks and UAV navigation, respectively; language modeling, where Liu et al. \cite{liu2025quantum}, Ni et al. \cite{ni2026quloc}, and Aizpurua et al. \cite{aizpurua2026quantum} report perplexity; object detection \cite{roh2024fast, kim2025quantum, roh2025hybrid}; point cloud upsampling \cite{zhu2025qctkd}; and combinatorial optimization for QAOA \cite{piperno2026knowledge}. Regression and representation learning remain largely unexplored in the context of quantum KD. Several studies acknowledge this limitation; for example, Yan et al. \cite{yan2026distilling} suggest extending classification KD to regression and representation learning; Li et al. \cite{li2025quantum} identify the application to generative tasks as a significant open challenge due to the limitations of qubit and circuit expressivity. Similarly, Chen et al. \cite{chen2026photonic} note that their framework has only been tested on basic image classification benchmarks. Therefore, expanding quantum KD to encompass a broader range of tasks is a natural progression and necessary to demonstrate its relevance beyond current experimental settings.

\section{Conclusion}
\label{sec:conclusion}
This paper presented a survey of quantum KD using a taxonomy that distinguishes three teacher-student configuration directions \textemdash classical-to-quantum (C2Q), quantum-to-quantum (Q2Q), quantum-to-classical (Q2C) \textemdash and quantum-enhanced KD, a complementary category defined by the use of quantum resources within the distillation process. The survey covers several applications, including image classification, object detection and 3D vision, reinforcement learning, natural language processing, combinatorial optimization, wireless communication, cybersecurity, and healthcare.

Classical-to-quantum KD is the most extensively studied direction in the reviewed literature, with multiple studies demonstrating improvements over non-distilled quantum baselines, although several report no benefit or degradation \cite{amin2025quantum, boucher2025n, nishan2026quantummedkd, innan2026towards}. However, hardware evaluation and comparison with classical counterparts under matched conditions remain limited.

Quantum-to-quantum KD has improved compressed students relative to comparable students trained without distillation, although retention of teacher accuracy depends on the compression setting. For example, Yan et al.’s seven-to-two-qubit experiment increases student accuracy from 52.3\% to 81.7\%, while the teacher achieves 98.3\% \cite{yan2026distilling}. In contrast, quantum-to-classical remains an emerging direction, with one study indicating that quantum-derived supervisory signals can improve the robustness of a deployable classical student under noise and rotation, and another study in which a quantum policy is distilled into a classical proxy for UAV control. Quantum-enhanced distillation embeds quantum kernels, photonic samplers, or gating circuits in classical distillation pipelines, but whether these components offer an advantage over classical alternatives, given their additional cost, remains an open question. Despite this progress, the field faces several unresolved challenges, including output mismatch between teacher and student, hardware constraints and noise, difficulty of attributing gains to the distillation, the dominance of response-based KD, and limited task diversity beyond classification.

Future work should focus on controlled standardized evaluations that isolate the effect of distillation, hardware-aware distillation methods, and more diverse distillation mechanisms, such as feature-based, relation-based, and circuit-level approaches, as well as further exploration of understudied configurations.

\bibliographystyle{IEEEtran}
\bibliography{citations} 

\end{document}